# Magnetic- and electric-field-induced anomalous Hall currents from optical excitation of Landau transitions in bulk GaAs

**Christoph Dresler, Shekhar Priyadarshi, and Mark Bieler***

*Physikalisch-Technische Bundesanstalt, Bundesallee 100, 38116 Braunschweig, Germany*

*Corresponding author: mark.bieler@ptb.de

We study a certain type of anomalous Hall current in magnetically- and electrically-biased bulk GaAs samples under the formation of Landau levels. The currents are generated by ultrafast optical excitation of spin-polarized carriers and detected by time-resolved measurements of the simultaneously emitted THz radiation. Due to the requirements of simultaneous magnetic and electric driving fields we refer to these currents as BE-AHC. We find that the BE-AHC peak for optical transitions between the band extrema of Landau levels of valance and conduction bands. These discrete features are attributed to the energy dependence of the geometric phases being responsible for anomalous transport effects in a semiconductor bandstructure. Surprisingly, we even detect the discrete Landau band transitions at room temperature, most likely due to the ultrafast local probing realized in our experiment. An analysis of the THz spectra using a model, based on the Boltzmann transport equation of optically excited carriers, shows that electron and hole contributions lead to complex current dynamics. While the cyclotronic motion of electrons results in a dip in the THz spectral response of BE-AHC, it causes a peak in the spectral response of normal Hall currents. Additionally, our experimental results strongly suggest that the Landau levels of the valance band play a significant role in the generation of BE-AHC. We expect that our results will initiate further studies on Berry-phase effects in Landau bands.

## I. Introduction

We start with a brief introduction to different variants of the Hall effect, being one of the most well-known physical effects in solid states [1]. This introduction is important since we distinguish between current components that have different requirements with respect to the carriers' spin polarization and electric and magnetic driving fields. The Hall effect denotes the deflection of charge carriers in an electrical conductor into a direction perpendicular to orthogonally applied electric and magnetic fields, see Fig. 1(a). The corresponding Hall current is obtained for both, electrically and optically injected carriers, does not depend on the carriers' spin polarization, and is proportional to the applied magnetic field $B$. In materials with broken time-inversion symmetry, such as ferromagnets, an additional contribution to the Hall resistivity has been found. This contribution, which was initially assumed to be proportional to the magnetization $M$ and independent of $B$, is known as anomalous Hall effect [2–8].

A similar effect, being closely related to the anomalous Hall effect, is observed in electrically biased materials *without* an externally applied magnetic field and *no* intrinsic spin polarization, but where time-inversion symmetry is broken by optical excitation of spin-polarized carriers. In this case, an anomalous current is obtained in the direction perpendicular to the electric driving field. The microscopic origins behind this effect involve the so called Berry curvature, an effect being intrinsic to the material's bandstructure, and extrinsic effects, such as side-jump and skew-scattering, that rely on scattering processes between the charge carriers and impurities and/or phonons [4,6,9–12]. Since the microscopic origins are the same as for the anomalous Hall effect, this photocurrent is also referred to as inverse spin-Hall effect or anomalous velocity [10,13,14], see Fig. 1(b).

Using again a material without intrinsic spin polarization, but with existing electric $\mathbf{E}$ *and* magnetic $\mathbf{B}$ fields, that are orthogonal to each other, another anomalous photocurrent exists [15–17]. This current flows in the direction of the applied magnetic field, see Fig. 1(c), and it requires dynamical equations, in which the time derivative of the momentum, $\dot{\mathbf{k}}$, is proportional to $(\mathbf{E} + \mathbf{v} \times \mathbf{B})$, with $\mathbf{v}$, being the carrier velocity. In the following, we refer to this special current component as magnetic- and electric-field induced anomalous Hall current (BE-AHC). *The abbreviation is chosen to clarify that the BE-AHC as defined by us is a special type of current, but belongs to the large family representing anomalous Hall effects.*

The first study on BE-AHC was performed on a bulk GaAs crystal being excited with circularly polarized light having a photon energy above the GaAs bandgap [15]. The BE-AHC generation occurred in two steps. (i) The optical excitation resulted in an electron spin current normal to the sample surface due to the diffusion of the spatial electron distribution. (ii) An external magnetic field was applied along a direction in the plane of the sample surface, which resulted in BE-AHCs flowing along the magnetic field direction. Since the currents were excited with a continuous-wave laser and detected via charge collection at electrodes, no temporal dynamics of the currents could be extracted. The initial electron spin current resulted from diffusion due to the gradient of the spatial electron distribution. The same effect will be obtained when replacing the diffusion current by a drift current (see paragraph below) and this is why we categorize the observations of Ref. [15] in the group of BE-AHCs, as defined above.

More recent studies on BE-AHCs in GaAs employed a very similar sample geometry [16,19]. Yet, the optical excitation was accomplished with femtosecond pulses, resulting in terahertz (THz) radiation emitted from the BE-AHC. For ultrafast excitation of GaAs the electric field that forms at the GaAs surface has been found to dominate over diffusion effects, i.e., Photo-Dember effect [18]. Thus, the electric surface field was assumed to be the main contributor for the initial acceleration of the optically induced carriers. Again, an external magnetic field was applied along a direction in the plane of the sample surface, resulting in BE-AHCs flowing along the magnetic field direction. While the THz peak-to-peak signal emitted from the BE-AHCs was used to extract their strength, the THz shape gave access to BE-AHC dynamics on a sub-picosecond time scale. However, these studies were limited to room temperature and weak magnetic fields ($\leq$ 1.2 T).

In this paper, being motivated by the so far unstudied influence of Landau levels on BE-AHCs, we extend the previous THz investigations of Refs. [16,19] to cryogenic temperatures down to 10 K and strong magnetic fields up to 7 T. Under these conditions we observe a discretization of BE-AHCs corresponding to optical transitions between valance and conduction Landau bands, being revealed at different photon energies of the excitation pulses and magnetic field strengths. Surprisingly, we find that the presence of Landau features in BE-AHC measurements is not limited to cryogenic temperatures but appears in room temperature experiments, too. These observations endorse the sensitivity of THz-based studies of ultrafast photocurrents. Analyzing the shape of THz radiation from BE-AHCs and comparing them to a model of BE-AHCs based on the Boltzmann equations, we identify differences in time and frequency domain between the BE-AHC contributions from the valance band (which follows the optical pulse envelope) and from the conduction band (which involves cyclotronic motion of electrons). Moreover, our results strongly suggest that the valance band contribution to the BE-AHC decreases for increasing magnetic fields.

The remainder of this paper is structured as follows. In Sec. II, we introduce the samples and the experimental setup, including a brief description of the BE-AHC generation process. We comment on the processing of the measured THz traces for the extraction of BE-AHCs in Sec. III. The experimental results are discussed in Sec. IV and compared to a theoretical model based on the Boltzmann transport equations in Sec. V. This model is further detailed in the supplementary material [20]. Finally, the paper is concluded in Sec. VI.

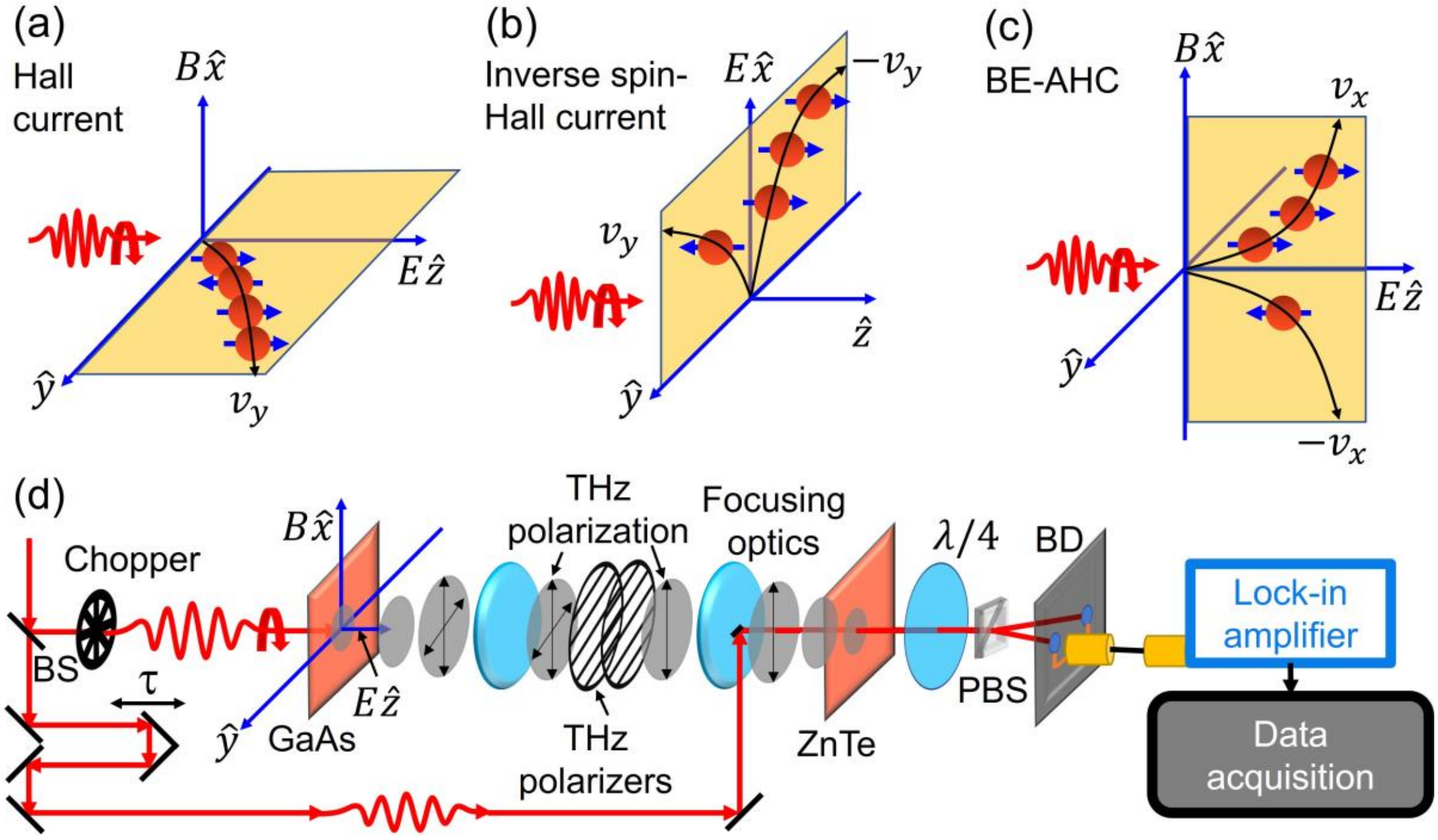

Fig. 1: Variants of the Hall current. (a) Hall current under applied electric and magnetic field. (b) Inverse spin Hall current in a material without intrinsic spin polarization resulting from an electric driving field only. (c) Magnetic- and electric-field-induced anomalous Hall current (BE-AHC) in a material without intrinsic spin polarization resulting from orthogonally-oriented electric and magnetic driving fields. (d) Experimental scheme of the pump-probe setup. A pulsed laser source is split into pump and probe pulses at a beam splitter (BS). The circularly-polarized pump pulses excite a GaAs sample held in an optical cryostat. The excitation results in BE-AHC and HC generation in the presence of a surface electric field $E$ and an external in-plane magnetic field $B$. The simultaneously emitted THz radiation is collected from the backside of the sample and detected via electro-optic sampling in a ZnTe crystal employing the linearly-polarized probe pulses. Electro-optic detection involves an ellipsometric analysis of the probe beam polarization using a quarter-waveplate (λ/4), a polarizing beam splitter (PBS), and a balanced detector (BD). The output of the BD is detected with a lock-in amplifier at the frequency at which the optical chopper modulates the pump beam. A delay stage is used to vary the temporal delay $\tau$ between the pump and probe pulses allowing for time-equivalent sampling of the temporal shape of the THz traces. The experiments are done at either room temperature or 10 K.

## II. Experimental methods

In this section we introduce the experimental setup, whose main part is shown in Fig. 1(d), and comment on the basics of our generation schemes for HC and BE-AHC. The experiments were performed on 350 μm thick semi-insulating GaAs bulk samples with crystallographic (001) and (110) orientations. The sample under study was placed inside a magneto-optical cryostat, where it was subjected to a magnetic field along an in-plane direction of the sample. The magnetic field could be continuously varied between 0 and ±7 T. For optical excitation of the samples at normal incidence, circularly polarized pump pulses with a pulse width of 140 fs, a repetition rate of 76 MHz, and a peak intensity of 10 MW/cm$^2$ were obtained from a Ti:Sa laser, whose center wavelength can be tuned. The wavelength tuning was used to realize different photon excitation energies with respect to the bandgap of GaAs. We note that due continuous laser heating we assume a cryogenic base temperature of the samples of 10 K rather than the 4.2 K provided by the cryostat.

The optical excitation resulted in a sub-picosecond generation of currents. This process leads to the emission of THz radiation. The THz radiation was collected from the backside of the samples and detected in a free-space manner employing a standard electro-optic (EO) sampling technique [21,22], see Fig. 1(d). The optical sampling (probe) pulses were obtained from the same Ti:Sa laser, from which the pump pulses were obtained. The electric-THz field induced a birefringence in the EO crystal (1.5 mm thick ZnTe), resulting in a change of polarization of the probe pulses, which, in turn, was measured using an ellipsometric technique [22]. A mechanical delay line was used to change the time delay between the THz and probe pulses, allowing for time-equivalent sampling of the shape of the THz traces emitted from the optically induced currents in the GaAs samples. Since in the far field the THz polarization is parallel to the direction of current flow, we used two wire-grid THz polarizers to only detect current components flowing in a certain in-plane direction of the sample. The suppression of unwanted current components was further improved by the THz polarization dependence of the ZnTe crystal. Together we estimate a suppression of THz signals from currents flowing in an unwanted direction of approximately 400:1 [19] as compared to THz signals from currents flowing in the direction of study.

We now comment on the generation schemes for the BE-AHC and HC. As indicated in Fig. 1(a) and (c), both currents are induced in the sample plane and are orthogonally oriented with respect to each other such that orthogonally polarized THz radiation propagating in the direction normal to the samples' surface will be emitted. The coordinate system of the experimental setup is denoted by $(\hat{x}, \hat{y}, \hat{z})$. Different coordinate systems $(\hat{x}_{110}, \hat{y}_{110}, \hat{z}_{110})$ and $(\hat{x}_{001}, \hat{y}_{001}, \hat{z}_{001})$ are used for the crystallographic directions of the samples, see below. For the generation of AHCs we employ circularly polarized laser pulses. The created carriers acquire a velocity perpendicular to the surface due to $E\hat{z}$ that forms at the GaAs surface. The static magnetic field $B\hat{x}$ then rotates the spin and direction of current flow due to Larmor precession and cyclotron rotation, respectively. This creates the so-called spin-Hall angle resulting in an BE-AHC parallel to the applied magnetic field along the $\hat{x}$ direction. Here we would like to note two things. (i) Following Ref. [18] we assume that the surface electric field dominates over the Photo-Dember effect in GaAs. With an absorption coefficient of $1.2 \times 10^4$ cm$^{-1}$, approximately 50% of the carriers will be absorbed within an electric-depletion-field region of 500 nm [18]. Yet, even if the Photo-Dember effect was mainly responsible for the initial spin current flow, the conclusions of this work would remain to be valid. (ii) The Larmor frequency of electrons is approximately 4 GHz at $B = 1$ T and about a factor of 100 smaller than the corresponding cyclotron frequency [16,19]. Thus, we can safely assume that for the THz experiment done in this study, cyclotron rotation dominates over Lamor precession.

The BE-AHC flows in opposite direction upon inversion of the spin orientation, i.e., for changing the helicity of the optical excitation. For the generation of HCs the sample is again excited with circularly polarized light creating a carrier population with a net spin orientation perpendicular to the surface. Similar to the BE-AHC generation, the carriers are accelerated in the intrinsic electric field $E\hat{z}$ and the applied magnetic field $B\hat{x}$ induces a Lorentz force yielding HCs along the $\hat{y}$ direction. Here we like to emphasize that, in contrast to the BE-AHC, the HC does not depend on the spin of the carriers, i.e., on the helicity of the optical excitation, which has been experimentally verified. Moreover, the same results for HC generation are obtained for excitation with linear polarization, which leads to an equal amount of spin-up and spin-down carriers. In the actual experiments on HCs we employed circularly polarized light for the generation of HCs to ensure that HCs and BE-AHCs are generated and measured under exactly the same experimental conditions. This increases the confidence in our comparative study of HCs and BE-AHCs.

## III. Extraction of BE-AHCs from measured THz traces

Despite the aforementioned THz polarization sensitivity of the detection setup, we needed to undertake additional measures for the exclusive detection of BE-AHCs and this is described in this section. This necessity mainly arises from the occurrence of additional current components and from imperfections of the experimental setup. In the following we distinguish between the two different orientations of the GaAs samples.

*(110)-orientation:* There are two strong in-plane photocurrent components coexisting with BE-AHCs in (110)-oriented GaAs bulk samples for normal optical excitation, namely shift currents and HCs. The shift current results from a spatial shift of the center of the electron charge during optical excitation [21,23–27]. We define the crystallographic directions in the (110)-oriented GaAs bulk sample as $\hat{x}_{110} = [001]$, $\hat{y}_{110} = [1\bar{1}0]$, and $\hat{z}_{110} = [110]$. To induce BE-AHCs along $\hat{x}_{110}$, a magnetic field $B$ is applied along $\hat{x}_{110}$ and the sample is excited with circularly-polarized optical pulses at normal ($\hat{z}_{110}$) incidence. To analyze the possible generation of shift currents we express the circular polarization with two (phase-delayed) polarization components oriented along the $\hat{x}_{110}$ and $\hat{y}_{110}$ directions. The $\hat{y}_{110}$ component induces a shift current directed along the $\hat{x}_{110}$ direction employing the xyy shift current component in (110)-oriented bulk GaAs [28,29]. This needs to be accounted for in the BE-AHC measurements. (The $\hat{x}_{110}$ component of the circularly-polarized light does not induce any shift current as corresponding tensor elements do not exist.) Additionally, as described in Sec. II, the surface field $E\hat{z} = E\hat{z}_{110}$ and the applied magnetic field $B\hat{x} = B\hat{x}_{110}$ lead to HCs along the $\hat{y} = \hat{y}_{110}$ direction. Normally, this current would be suppressed by the THz polarizers. However, despite the suppression factor of 400:1 of the EO detection setup, no complete suppression of the HCs was obtained. Therefore, we have employed the following symmetry relations to eliminate both, shift and Hall currents. The BE-AHC flips its sign with the direction of the applied magnetic field $(-B\hat{x}, +B\hat{x})$ and helicity of the circular polarization ($\sigma^+$ and $\sigma^-$ for right- and left-handed circular polarizations, respectively). The shift current does not flip its sign with either $B$ or $\sigma$. The HC flips its sign with $B$ but not with $\sigma$, since it is spin independent. Consequently, we measured THz traces $E_{\text{THz}}$ from currents flowing along the $\hat{x}$ direction for four different excitation conditions with respect to $B$ and $\sigma$: (i) $[+B\hat{x}, \sigma^+]$, (ii) $[+B\hat{x}, \sigma^-]$, (iii) $[-B\hat{x}, \sigma^+]$, and (iv) $[-B\hat{x}, \sigma^-]$. Afterwards we extract a THz difference signal from:

$$E_{\text{diff}} = (E_{\text{THz}}[+B\hat{x}, \sigma^+] - E_{\text{THz}}[-B\hat{x}, \sigma^+] - E_{\text{THz}}[+B\hat{x}, \sigma^-] + E_{\text{THz}}[-B\hat{x}, \sigma^-])/4. \qquad (1)$$

In the (110)-oriented samples this difference signal corresponds to the THz trace emitted from the BE-AHC along $\hat{x} = \hat{x}_{110}$ [16]. This difference technique also helps to suppress other current components, being $B$ and/or σ independent, due to non-perfect experimental geometries.

*(001)-orientation:* In (001)-oriented GaAs bulk samples, the extraction of BE-AHC components becomes even more complicated. First, we set the crystallographic directions as $\hat{x}_{001} = [110]$, $\hat{y}_{001} = [1\bar{1}0]$, and $\hat{z}_{001} = [001]$. In this geometry, there exists no shift current for optical excitation at normal incidence. However, along with the BE-AHC and HC there exists a current component along $\hat{x}_{001}$, referred to as circular bulk magneto photocurrents (CBMC), which flips its sign with both, inversion of the magnetic field $B\hat{x} = B\hat{x}_{001}$ and inversion of the helicity σ and is independent of the surface field $E\hat{z} = E\hat{z}_{001}$ [17,19,30,31]. The CBMC is a photo-galvanic effect, which appears in a non-centrosymmetric bulk crystal upon excitation with circularly-polarized light and in the presence of a magnetic field perpendicular to the propagation direction of the light beam. Ivchenko *et al.* [31] theoretically showed that the CBMC results from $k$-cubic terms in the conduction band Hamiltonian. Yet, the dependence of the CBMC on rotation of the sample by 90 degrees is different as compared to the BE-AHC. If the CBMC flows along $+\hat{x}_{001}$ for the magnetic field applied along $+B\hat{x}_{001}$ and for a certain helicity, it will flow along $-\hat{y}_{001}$ for $+B\hat{y}_{001}$ upon rotation of the sample by 90 degrees while keeping the optical helicity fixed. This symmetry relation changes for the BE-AHC. If the BE-AHC flows along $+\hat{x}_{001}$ for the magnetic field applied along $+B\hat{x}_{001}$ and for a certain helicity, it will flow along $+\hat{y}_{001}$ for $+B\hat{y}_{001}$ upon rotation of the sample by 90 degrees while keeping the optical helicity fixed. Therefore, to extract the BE-AHC in experiments on the (001)-oriented GaAs bulk sample, THz experiments had to be performed for two rotation directions of the sample ($\hat{x}_{001}$ parallel to $B\hat{x}$ and $\hat{y}_{001}$ parallel to $B\hat{x}$) with the other experimental conditions being identical. For each of the two rotation directions of the sample, the difference signal $E_{\mathrm{diff}}$ was calculated according to Eq. (1). Finally, the average of the two difference signals $E_{\mathrm{diff}}(\hat{x}_{001}||B)$ and $E_{\mathrm{diff}}(\hat{y}_{001}||B)$ was taken to obtain the BE-AHC in the (001)-oriented samples, thereby eliminating the CBMC.

Exemplarily we now show a step-by-step extraction of the THz signal emitted from the BE-AHC in the (001)-oriented GaAs sample for an excitation photon energy of 40 meV above the GaAs bandgap [32], an applied magnetic field of 7 T, and a temperature of 10 K. In Fig. 2a are plotted four THz traces obtained with the experimental parameters $[+B\hat{x}_{001}, \sigma^{+}]$, $[+B\hat{x}_{001}, \sigma^{-}]$, $[-B\hat{x}_{001}, \sigma^{+}]$, and $[-B\hat{x}_{001}, \sigma^{-}]$. Fig. 2(b) shows the four THz traces obtained after rotation of the sample by 90 degrees, i.e., $[+B\hat{y}_{001}, \sigma^{+}]$, $[+B\hat{y}_{001}, \sigma^{-}]$, $[-B\hat{y}_{001}, \sigma^{+}]$, and $[-B\hat{y}_{001}, \sigma^{-}]$. Using Eq. (1), the four THz traces of Fig. 2(a) and (b) lead to a difference signal $E_{\text{diff}}$ shown in Fig. 2(c) and (d), respectively. Finally, we take the average of the two THz traces plotted in Fig. 2(c) and (d) to obtain the THz trace emitted from the BE-AHC in the (001)-oriented sample, see Fig. 2(e). Here, we also note that the $\hat{x}_{001} = [110]$ and $\hat{y}_{001} = [1\bar{1}0]$ crystallographic directions are identical in the (001)-oriented GaAs sample, therefore the BE-AHCs along the two crystallographic directions are identical and the extracted BE-AHC represents the BE-AHCs along both directions. This is in contrast to the (110)-oriented GaAs sample in which we could only access the BE-AHC along the $\hat{x}_{110} = [001]$ crystallographic direction.

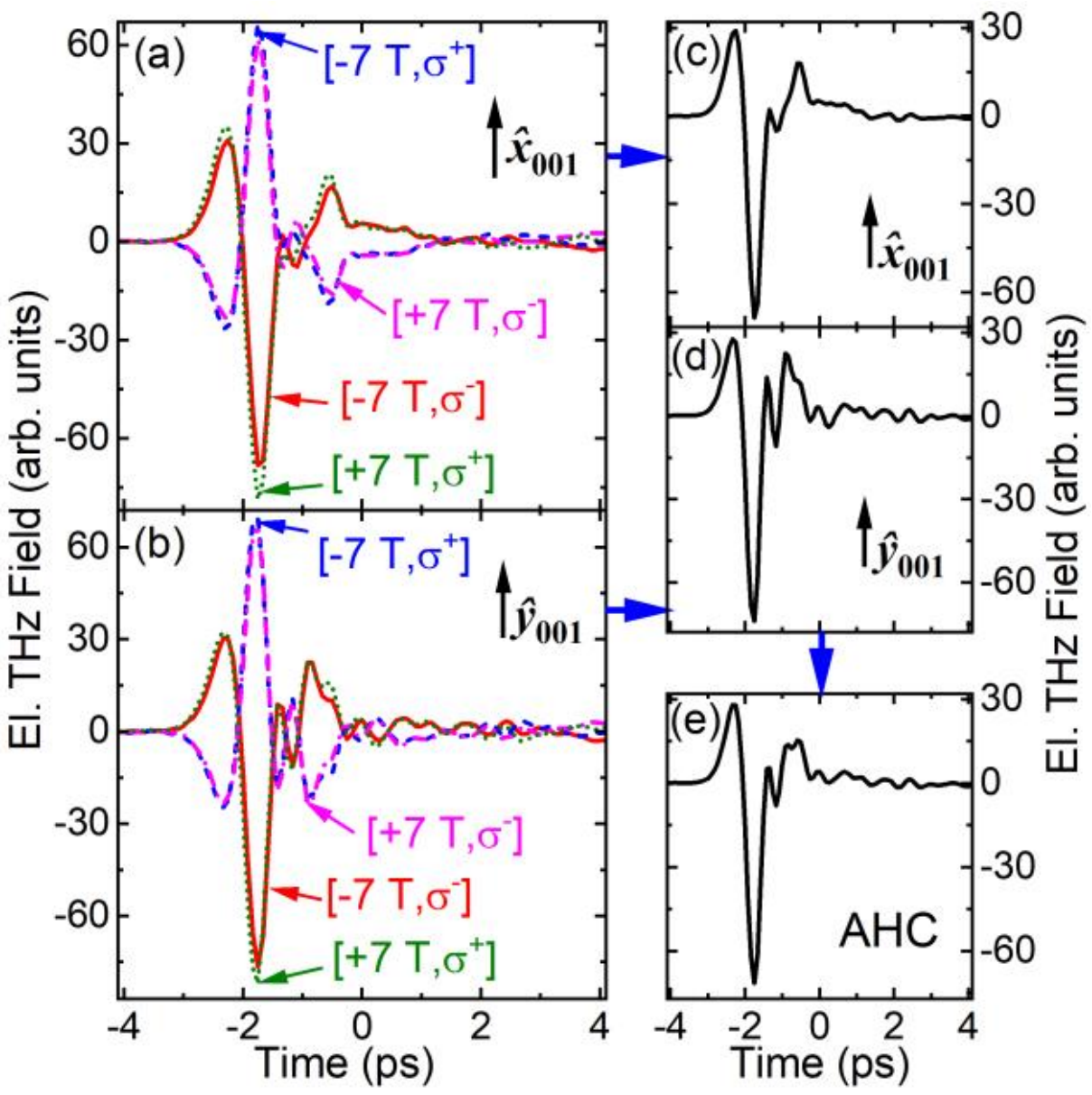

Fig. 2: Extraction of the BE-AHC in a (001)-oriented GaAs sample. THz traces for $[+7\ \text{T}, \sigma^{+}]$ (green dotted line), $[-7\ \text{T}, \sigma^{+},]$ (blue dashed line), $[+7\ \text{T}, \sigma^{-},]$ (purple dot-dashed line), and $[-7\ \text{T}, \sigma^{-}]$ (red solid line) for the magnetic field and the THz polarization oriented along $\hat{x}_{001}$ (a) and $\hat{y}_{001}$ (b). (c) and (d) THz traces resulting from the difference technique mentioned in the main text applied to the THz traces shown in (a) and (b), respectively. (e) Average of the THz traces plotted in (c) and (d) yielding the THz trace from the BE-AHC in the (001)-oriented GaAs sample. The signal-to-noise ratio of the individual THz traces shown in (a) and (b) is approximately 500.

Comparing the THz traces plotted in Fig. 2(a) with each other we obtain only small deviations from the expected symmetry relations that govern the BE-AHC generation. For example, the shapes of the THz traces linked to $[+B\hat{x}_{001}, \sigma^{+}]$ and $[-B\hat{x}_{001}, \sigma^{-}]$ are nearly identical and only differ slightly in amplitude. The same applies for the THz traces plotted in Fig. 2(b). This result endorses the excellent suppression of unwanted current components in our experimental setup. Comparing the THz traces plotted in Fig. 2(c) and (d), we observe a difference in the shape of the THz traces and in amplitude. This confirms the coexistence of BE-AHC and CBMC. Taking the average of the two THz traces, plotted in Fig. 2(e), we obtain a THz trace originating solely from BE-AHC. Finally, it should be noted that the symmetry relations described in this section only needed to be taken into account for the study of BE-AHCs but not for the study of HCs, as the latter were much larger in amplitude compared to the BE-AHC.

## IV. Experimental results

We first present BE-AHC measurements under different experimental conditions. In Figs. 3(a), (b), and (c), are shown contour plots of the THz peak-to-peak signal emitted from BE-AHCs versus magnetic field (horizontal axis) and photon excitation energy with respect to the bandgap of GaAs (vertical axis). As mentioned above, the THz peak-to-peak signal emitted from BE-AHC mainly denotes the corresponding current strength. Fig. 3(a) and (b) were obtained on a (110)-oriented GaAs with the BE-AHC flowing along $\hat{x}_{110}$ at 10 K and at room temperature, respectively, whereas Fig. 3(c) was obtained on a (001)-oriented sample at 10 K. For both sample orientations, the BE-AHCs were extracted as detailed in Sec. III. In all three plots, the BE-AHC amplitude shows clear peaks whose positions depend on both excitation photon energy and magnetic field. To analyze the origin of these features we have also plotted in Figs. 3(a), (b), and (c), the bandgap energies required for optical transitions from the n=1, 2, and 3 heavy-hole Landau band to the corresponding conduction Landau bands. The values were obtained from Ref. [33], where magneto-optical transitions between valence and conduction band Landau levels have been studied. The authors used a thin and strained GaAs sample, with the strain leading to a shift of the involved transitions. To account for this, we shifted the energy scale by $+4$ meV to match our experiment and only use the transitions originating from heavy holes and neglected light holes. The neglection of light-holes is motivated due to the smaller dipole moment for light-hole to electron transitions as compared to heavy-hole to electron transitions. Moreover, the energy differences between light- and heavy-hole resonances in bulk

GaAs are very small and consequently a differentiation between the light and heavy hole transitions for pulsed femtosecond excitation is very difficult.

With the black lines an enhancement of the BE-AHC amplitude at Landau transition energies is immediately apparent and exists for both sample orientations, see Figs. 3(a) and (c). Surprisingly such enhancement of the THz peak-to-peak signal at the Landau transition energies also appears at 300 K, see Fig. 3(b). The direct correlation between BE-AHC enhancement and Landau transitions becomes obvious when studying HC measurements in (001)-oriented GaAs at 10 K, see Fig. 3(d). In order to detect THz emission from Hall currents, the THz polarizers and the EO sampling crystal have been rotated by 90 degrees, such that the EO detection is sensitive to in-plane current components flowing perpendicular to the applied magnetic field. In the HC measurements, no current enhancement at Landau transitions is observed. Instead, the HC simply peaks at the magnetic field at which the frequency spectrum emitted from the HC matches the spectral response of the EO detection. At a magnetic field of 3 T, the cyclotron frequency of electrons is approximately 1.33 THz, agreeing well with the maximum of the EO detection function [34]. Such a feature, resulting from the THz detection set-up, can also be appreciated in Figs. 3(a), (b), and (c) being superimposed on the Landau features. The ratio between HC and BE-AHC at a magnetic field of 3 T and an excitation photon energy of 10 meV above the bandgap of GaAs is approximately 30 and is also slightly depending on the orientation of the GaAs sample, compare the color scales in Fig. 3. In any case, the missing current enhancement at the Landau transitions in the HC measurements directly proves that the current enhancements at the Landau transitions seen in Figs. 3(a), (b), and (c) are directly linked to the BE-AHC.

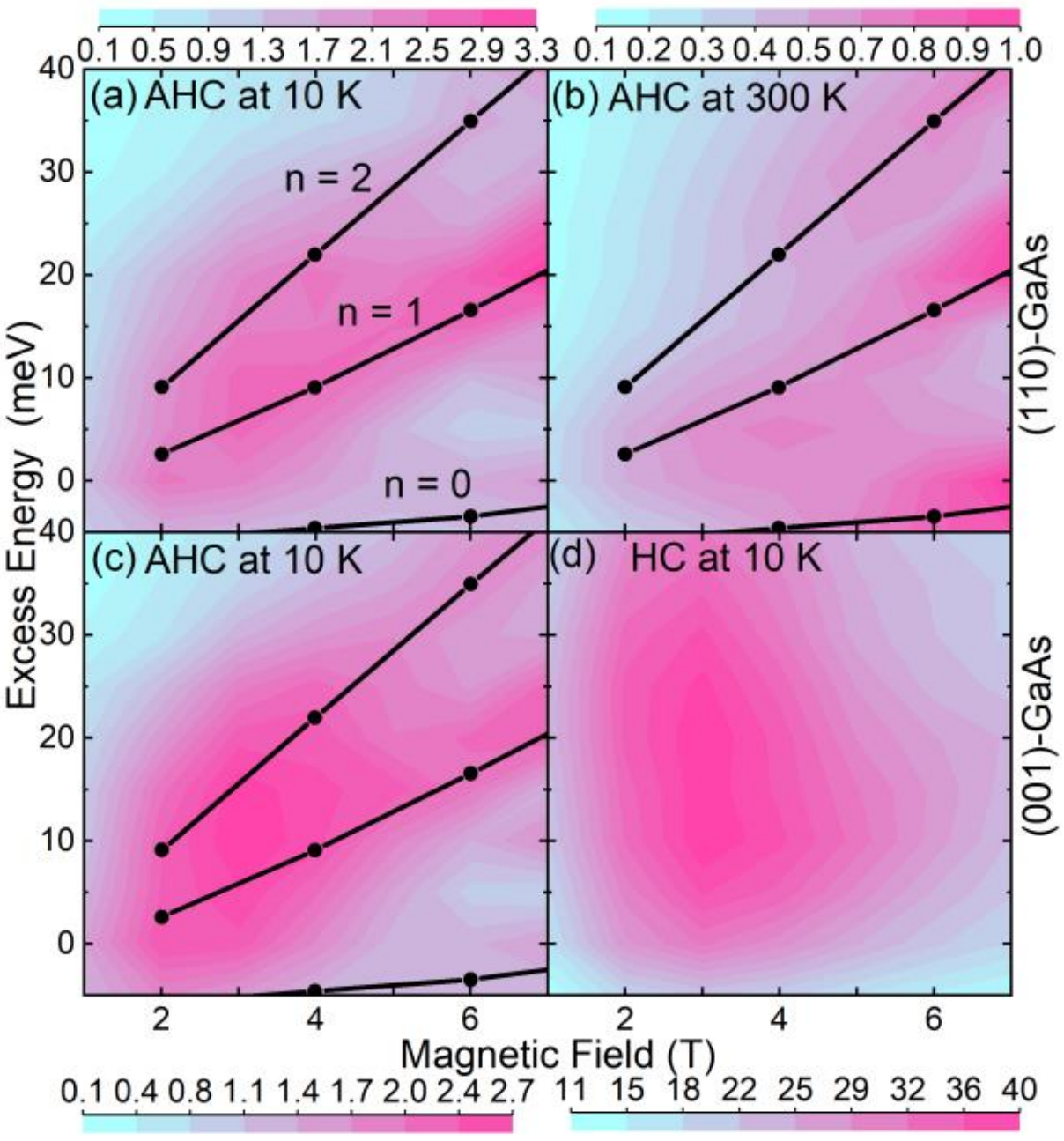

Fig. 3: THz peak-to-peak signal from the BE-AHC and HC in two different samples: The top row shows BE-AHCs along $\hat{x}_{110}$ in the (110)-oriented GaAs sample at 10 K (a) and 300 K (b). The bottom row shows the results for the (001)-oriented GaAs sample at 10 K for the BE-AHC (c) and the HC along $\hat{y}$ (d). The amplitude of the contour plot of (b) is normalized to one and the amplitude of all other plots are scaled by the same normalization factor, i.e., all amplitudes are directly comparable to each other. In (a), (b), and (c) are shown photon energies required for the transition between $n^{th}$ conduction and valance Landau bands (black lines).

To understand the physical origin of the BE-AHC enhancement at the Landau transition energies, it is necessary to discuss the BE-AHC generation process in more detail. In the presence of an external magnetic field the carriers in a bulk semiconductor are confined to hollow cylinders due to cyclotronic motion, leading to a reduction of the density of state from 3 D to 2 D [35]. With such a reduction of the density of states we would expect to see a step like enhancement of the THz peak-to-peak signal versus photon excess energy in the contour plots. Instead we observe discrete-like features of the THz peak-to-peak signal at the Landau transition energies, corresponding rather to a 1 D instead of a 2 D density of states. We believe that the photon energy dependence of the BE-AHC is largely influenced by the band energy dependence of the microscopic origins of the BE-AHC and that this is the main reason for the discrete features seen in the BE-AHC measurements.

The microscopic origins of the AHC can be divided into intrinsic effects (Berry phase/curvature) and extrinsic effects (side-jump and skew scattering), with the latter being linked to the so-called Pancharatnam phase [6]. Both, Berry phase and Pancharatnam phase are geometric phases. In an unbiased semiconductor bandstructure these geometric phases are stronger for states close to the bandgap as compared to states at large excess energies [4,6,9,11]. We anticipate that such dependence of the geometric phases also holds for the Landau bands and, consequently, leads to discrete features in the BE-AHC measurements. Two other factors might contribute to the enhancement of BE-AHCs at the Landau transitions. First, mixing between light- and heavy-holes in the valence band is known to have significant influence on current dynamics in lower dimensional semiconductors [36]. (In our case, it is not the quantum well, but the magnetic field that leads to a 2D density of states, as mentioned above.) In such structures, bandmixing is often strongest close to the band extrema and decreases when going higher into the band [36–38]. Since geometric phases are known to be stronger for stronger bandmixing [9], we expect to obtain stronger BE-AHCs for band-edge optical excitation. Second, in a recent study, it has been shown that the microscopic origins, that are also responsible for the BE-AHC generation, are stronger for exciton excitations [11]. This also makes excitonic excitations at the Landau band edges a possible candidate for the enhancement of AHCs observed in our measurements. In any case, it is very likely that the complexity of the bandstructure, which manifests itself in bandmixing and excitonic states, is responsible for BE-AHC enhancements. In contrast, the HC does not depend on the geometric phases and therefore, it does not inherit the discrete features seen in Figs. 3(a), (b), and (c). A detailed discussion on the BE-AHC and HC dynamics will be given in Sec. V.

As discussed above the discrete Landau features in the BE-AHC not only appeared at 10 K, but also at 300 K, see Fig. 3(b). In our experimental geometry we detect the THz radiation emitted from the BE-AHC. Since the THz radiation corresponds to the time derivative of the current, fast current changes such as the onset of the current flow will be detected. In other words, due to the probing on ultrafast time scales, the THz studies of photocurrents are partly immune to scattering effects leading to transition broadening in semiconductors. Thus, the spectral resolution in probing discrete transitions is mainly limited by the spectral width of the femtosecond laser, which is approximately 6 nm in our work. This detection is an example of probing transport effect at their spatial origin, which eliminates the need for charge collection at electrodes. We believe that this local probing is the reason for the observation of the Landau signatures in the BE-AHC measurements even at room temperature. In the room temperature measurements we observe qualitatively the same features as compared to 10 K, yet, with a lower THz peak-to-peak signal, see Figs. 3(a) and (b). We note that another example of such local probing of currents was presented in Ref. [39], where the authors employed Faraday rotation to obtain THz response from currents in the quantum Hall regime, yet, with sample temperatures well below 10 K.

So far, we only commented to the peak-to-peak signal of the THz traces emitted from the BE-AHC. However, the THz measurements also allow us to study the temporal dynamics of the BE-AHC. In Fig. 4 are plotted THz traces measured at the four corners of the contour plot of Fig. 3(a), i.e., at [1 T, +40 meV] in (a), [7 T, +40 meV] in (b), [1 T, -5 meV] in (c), and [7 T, -5 meV] in (d). Analyzing the shape of the THz traces (see also red vertical lines, denoting the same temporal instance for all four THz traces) we surprisingly observe that the BE-AHC [1 T, -5 meV], see Fig. 3(c), flows opposite to the BE-AHC resulting from the other excitation conditions. At present we can only speculate about the reason for the current inversion. This behavior resembles to some extent the photon excitation energy dependence of the so-called shift and injection currents studied in Refs. [36,38,40]. In those studies, the photocurrents obtained from for the optical excitation of excitons transitions were oppositely oriented as compared to the photocurrents obtained from excitation of continuum transitions. Valance bandmixing could be identified as the main reason for the current inversion. It might be that valence bandmixing is also responsible for the current inversion observed in Fig. 4(c).

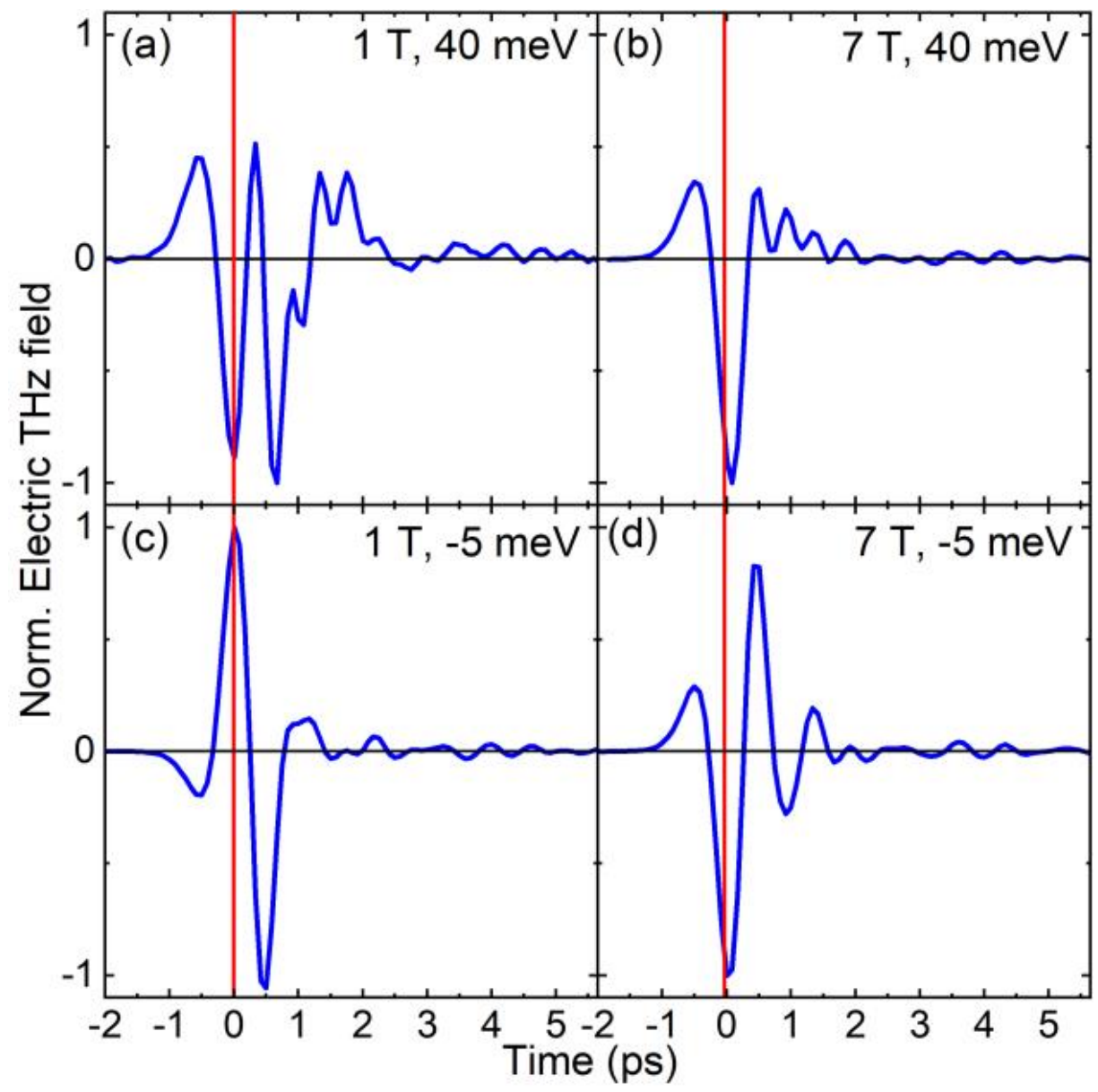

Fig. 4: Measured THz traces from BE-AHCs in the (110)-oriented GaAs sample held at 10 K for magnetic fields and excitation photon energies of [1 T, +40 meV] (a), [7 T, +40 meV] (b), [1 T, -5 meV] (c), and [7 T, -5 meV] (d). The energies are given relative to the calculated bandgap of bulk GaAs. The red arrows are guides to the eye. They point to the same temporal instance and the sign of the THz signal at this temporal instance can be taken as an indication of the BE-AHC direction.

Another observation which can be made in Fig. 4(a) is that the THz signal emitted from BE-AHCs at [1 T, +40 meV] shows an oscillatory behavior which does not occur in the other THz signals of Fig. 4. To understand this better it is helpful to look at the frequency domain representations of BE-AHCs and HCs. In the top panel of Fig. 5 we have plotted the THz spectra from BE-AHCs (solid red lines) measured in the (001)-oriented GaAs sample for a photon excess energy of 35 meV, a temperature of 10 K, and for different magnetic fields. Similar plots are shown in the bottom panel of Fig. 5 for the HC (solid red lines). The shape of the THz spectra emitted from BE-AHC and HC differ significantly for weak magnetic fields (< 4 T) while it tends to appear similar at stronger magnetic fields (> 4 T). For $B$ = 1, 2, and 3 T, we observe dips in the spectra of the BE-AHCs being slightly blue shifted as compared to the electron's cyclotron frequencies which is 1.33 THz at 3 T. With this we can relate the time-domain oscillations in Fig. 4(a) to the dips seen in the BE-AHC spectra at low magnetic fields in Fig. 5. More information on this relation will be given in Sec. V. In contrast to the spectral dips of the BE-AHC, the HC has spectral peaks being slightly red shifted as compared to the electron's cyclotron frequencies for $B = 1$, 2, and 3 T. Yet, this is not any more visible for $B > 3$ T, since at these magnetic fields the electron's cyclotron frequency exceeds the spectral detection range of the EO sampling set-up, supporting our previous statement on the influence of the frequency-limited EO transfer function on Fig. 3(d). We believe that this frequency-limited EO transfer function is also the reason why the corresponding THz spectra of the BE-AHC and HC are very similar at large magnetic fields.

## V. Comparison to simulations

To understand the spectral and time-domain features of the THz radiation emitted from BE-AHCs and HCs, we have setup a simple model based on the Boltzmann transport equations for optically excited carriers [6,10,41,42]. A full mathematical treatment is presented in the supplementary material [20]. Here, we only outline the results of the model and start with the BE-AHC. As mentioned above, there are three main microscopic origins of the BE-AHC: (i) Berry curvature, (ii) side-jump, and (iii) skew scattering [7,12,43]. The Berry curvature is an intrinsic property of the bandstructure and is spin dependent [4,44]. It causes an acceleration of charge carriers perpendicular to an external electric field and parallel to an external magnetic field. (The latter is true if carrier velocity and Berry curvature point into the same direction [20].) The side-jump is a real-space displacement of carriers that occurs during scattering processes with impurities or phonons [7,45]. The skew scattering represents an asymmetric scattering event in momentum space and consequently results in asymmetric carrier distribution in momentum space [7]. For the generation of BE-AHCs, an energy dependent skew scattering rate is required [46]. Since this is typically considered very weak, we ignore skew scattering in our model. The insignificance of the skew scattering processes as compared to the Berry curvature and the side-jump processes was also reported in Ref. [10]. From our model we obtain the following frequency responses of the Berry curvature, $J_{\mathrm{BC}}(\omega)$, and side-jump, $J_{\mathrm{sj}}(\omega)$, contributions to the BE-AHC:

$$J_{\mathrm{BC}}(\omega) \propto \frac{(i\omega+\gamma)I(\omega)}{((i\omega+\gamma)^2+\omega_B^2)(i\omega+w^{\mathrm{sf}})}, \quad [2]$$

$$J_{\mathrm{sj}}(\omega) \propto \frac{\gamma I(\omega)}{((i\omega+\gamma)^2+\omega_B^2)(i\omega+w^{\mathrm{sf}})}, \quad [3]$$

with $\gamma$ being the momentum scattering rate, $I(\omega)$ is the frequency domain optical pulse intensity, $\omega_B$ is the cyclotron frequency, and $w^{\mathrm{sf}}$ is the spin-flip scattering rate. To simulate the frequency response of the BE-AHC, that takes both, electron and hole contributions into account, we make the following approximations: (i) Conduction band (CB): We consider only side-jump contributions and neglect Berry curvature contributions for electrons in the CB, i.e., $J_{\mathrm{CB}}(\omega) = J_{\mathrm{CB,sj}}(\omega)$. This is because previous studies have shown that the side jump dominates [10]. (ii) Valence band (VB): We assume that the valance band contribution suffers extremely high scattering and becomes instantaneous, i.e., $J_{\mathrm{VB}}(\omega) = \big(J_{\mathrm{VB,BC}}(\omega) + J_{\mathrm{VB,sj}}(\omega)\big) \propto \frac{I(\omega)}{\gamma_{\mathrm{VB}} w_{\mathrm{VB}}^{\mathrm{sf}}}$. With these two approximations, we obtain the full THz spectral response from BE-AHCs including both, valence and conduction band contributions, as:

$$J_{\mathrm{AHC}}(\omega) = J_{\mathrm{VB}}(\omega) + J_{\mathrm{CB}}(\omega) \propto \left\{ \frac{\gamma_{\mathrm{CB}}}{\left((i\omega+\gamma_{\mathrm{CB}})^2+\omega_{B,\mathrm{CB}}^2\right)\left(i\omega+w_{\mathrm{CB}}^{\mathrm{sf}}\right)} + C \right\} I(\omega). \quad (4)$$

Here, $C$ is an arbitrary constant, which contains $\gamma_{\mathrm{VB}} w_{\mathrm{VB}}^{\mathrm{sf}}$ and accounts for the different strength of the CB and VB contributions. Because of bandmixing the contribution from the VB is expected to play an important role [9,10], and consequently, it must not be neglected.

For the HC our model yields a different spectral response. Since for the HC generation the Berry curvature and the side-jump do not play any role, the contribution from the VB is negligible as compared to the CB contribution and, consequently, we only consider the CB contribution to the HC. The HC is independent of the spin of carriers. In our model we assume that the HC decays with energy relaxation of electrons ($w^{\mathrm{E}}$). The reason for choosing $w^{\mathrm{E}}$ as a relaxation rate in the HC generation process, is the assumption that the HC is strong for hot electrons (carrying large momentum) and significantly decreases for cold electrons (carrying very small momentum). With this assumption and following the model detailed in the Supplementary Material [20], the THz frequency response of the HCs is expressed as:

$$J_{\mathrm{HC}}(\omega) = \frac{I(\omega)}{\left((i\omega+\gamma_{\mathrm{CB}})^2+\omega_{B,\mathrm{CB}}^2\right)\left(i\omega+w^{\mathrm{E}}\right)}. \quad (5)$$

For the simulation of the BE-AHC and HC, we set all the CB scattering and relaxation parameters (i.e., momentum scattering, spin-flip scattering, energy relaxation) to 2 ps$^{-1}$ [47] and additionally applied an appropriate EO transfer function [30] to convert the current spectra into THz spectra. With this approach $C$ is the only unknown variable. Since a microscopic derivation of $C$ is clearly out of the scope of this paper, we have to make some assumptions on $C$. We assume $C$ to be a positive number, i.e., the electron and hole current contributions flow into the same directions. There is no obvious and immediate reason why the contributions from electrons and holes should be counteracting. We then vary $C$ such that the simulation yields the best agreement to the experiment at $B = 1$ T, compare solid red and dashed blue lines in the upper left-hand side corner of Fig. 5. Here the dip in the measured spectrum is nicely reproduced by the simulations. Afterwards we repeat this procedure for the BE-AHC at magnetic fields of 2 T, 3 T, and 4 T This approach yielded a magnetic field variation of $C$ being very close to the function $C(B) = C(1\mathrm{T})\ \exp\{-(B/\mathrm{T} - 1)\}$ and we have used this function to extract the values for $C$ for the other magnetic fields. This decrease of $C$ for increasing magnetic field goes along with a decrease of the VB contribution to the BE-AHC for increasing magnetic field. Such a variation of the VB contributions is supported by the experimental data since at higher magnetic fields the BE-AHC and the HC have very similar spectra and the VB does not contribute to the HC.

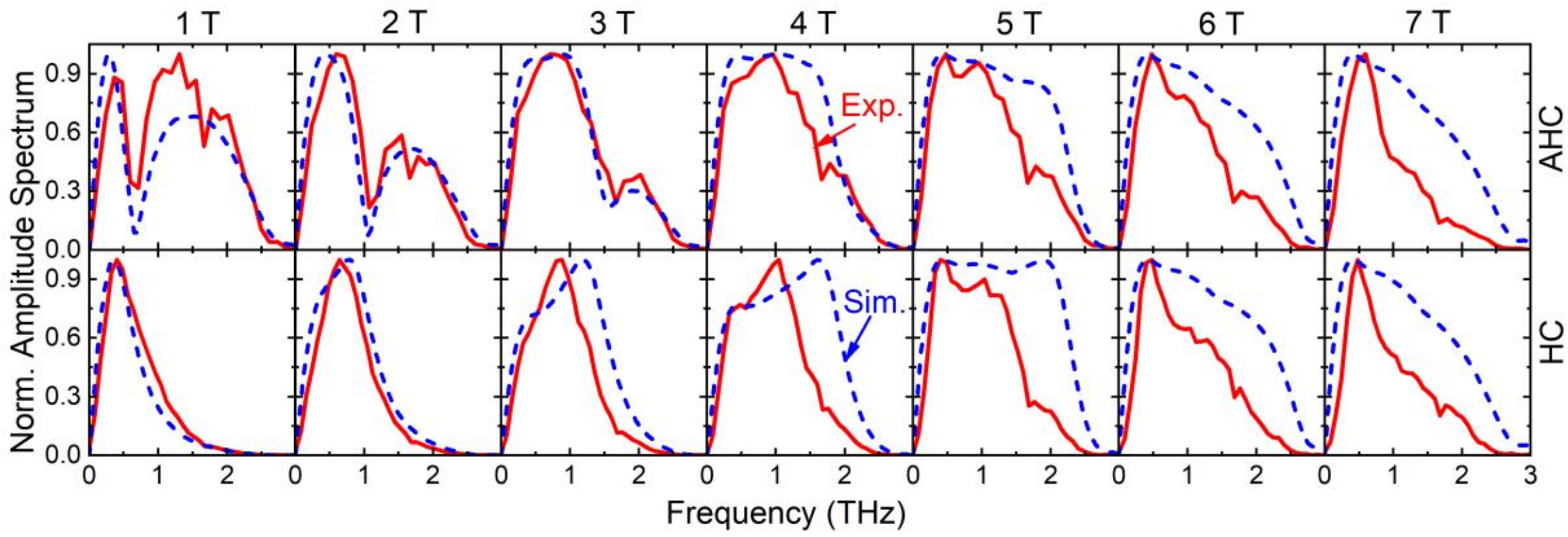

Fig. 5: Measured (solid red) and simulated (dashed blue) THz spectra emitted from the BE-AHC (top row) and HC (bottom row) for different magnetic fields ranging from 1 T to 7 T.

A plausible explanation for the decrease of the VB contribution to the BE-AHC at higher magnetic fields is valance bandmixing. In Refs. [9,10] the valance bandmixing was found to be responsible for the enhancement of the Berry curvature. We anticipate that the valance bandmixing will also enhance the side-jump and skew scattering processes. An increase in the magnetic field strength results in an increase in the separation of Landau bands. This may result in a reduction of the influence of valance bandmixing on Landau bands [48] and, in turn, a reduction of BE-AHCs evolving from Landau valence bands. This explains the required decrease of the value of $C$ with increasing the magnetic field strength. We like to point out that this observation resembles the observations presented in Ref. [10], where Berry curvature and side-jump related phenomena were stronger in wider quantum wells compared to those in narrower quantum wells. In wider quantum wells the separation between valance subbands is smaller compared to that in narrower quantum wells. The smaller this subband separation, the stronger the valance bandmixing and, consequently, the stronger the Berry curvature and side-jump phenomena [9]. The corresponding simulations of the BE-AHC are shown as blue dotted line in the upper panel of Fig. 5. Some deviations are obtained at higher magnetic fields. Yet, the dip occurring in the spectra at low magnetic fields is nicely reproduced. This dip results from an interference between CB and VB contributions to the BE-AHC and it is slightly blue shifted with respect to the electron's cyclotron frequency.

For the simulation of the HCs, the parameter $C$ does not have to be considered and the simulated THz spectra of the HC are shown in the bottom panel of Fig. 5 as blue dotted lines. As for the BE-AHC, some deviations are obtained at higher magnetic fields. However, the main features are qualitatively captured. The HC spectra show, even when correcting for the transfer function of the EO detection, a peak in the THz spectra that is slightly red shifted with respect to the electron's cyclotron frequency. The occurrence of dips (peaks) in the spectra of the BE-AHC (HC) manifests one of the main experimental and theoretical observations of our study. Finally, we note that decreasing the spin-flip scattering rate with increasing magnetic field may produce an even better resemblance between the experimental and simulated data for B > 3 T. However, owing to the uncertainty in our simulations we did not attempt to extract spin-flip scattering rates from the measurements.

## VI. Conclusions

In conclusion, we have observed discrete features related to optical interband Landau transitions in magnetic- and electric-field-induced anomalous Hall currents (BE-AHC). The THz-based detection technique enabled us not only to differentiate between the valance and conduction band contributions from the Landau bands to the BE-AHC but also allowed us to study the influence of valance bandmixing and exciton effects of Landau bands on the currents. Together with the experiments we have developed a simple model qualitatively explaining most of the observed dynamical properties of the BE-AHC. We like to note that a microscopic treatment of geometric phases and the description of the dynamical responses using, e.g., the semiconductor Bloch equations is very complicated and out of scope of our experimental paper. Yet, we hope that our initial work will trigger further experimental and theoretical studies. Since our work helps to distinguish between extrinsic and intrinsic effects occurring in Landau levels of valence and conduction bands, it might also be very useful to further engineer anomalous Hall effects in different types of semiconductors with the aim to observe quantum variants of the anomalous Hall currents even at room temperature.

**Acknowledgements:** This work was supported by the Deutsche Forschungsgemeinschaft (DFG) through the project BI 1348/2-1.

**Author contributions**: MB initiated the study. CD performed the experiments. SP developed the theoretical model. All authors discussed the results. MB and SP wrote the paper.

**Competing Interests:** The authors declare that they have no competing interests.

**Data availability:** The data that support the findings of this study can be obtained from the corresponding author upon reasonable request.

# Supplementary Material:
# Magnetic- and electric-field-induced anomalous Hall currents from optical excitation of Landau transitions in bulk GaAs

**Christoph Dresler, Shekhar Priyadarshi, and Mark Bieler***

*Physikalisch-Technische Bundesanstalt, Bundesallee 100, 38116 Braunschweig, Germany*

*Corresponding author: mark.bieler@ptb.de

In this Supplementary Material we present a mathematical model to explain the THz spectral features and current dynamics observed in magnetic- and electric-field-induced anomalous Hall current (BE-AHC) and Hall current (HC) studies as detailed in the main article. After comparing the following model with experiments, two main conclusions are obtained: (i) Both, valance (VB) and conduction (CB) bands contribute to the BE-AHCs and the CB cyclotron motion leads to a dip in the corresponding THz spectra emitted from BE-AHCs. This spectral dip cannot be explained without a significant contribution from the VB. (ii) The HCs are dominated by CB contributions and the corresponding THz spectra have peaks corresponding to the CB cyclotron frequency.

## I. Introduction to BE-AHC and HC

We start this Supplementary Material by describing a scheme for the generation of magnetic- and electric-field-induced anomalous Hall currents (BE-AHCs) and Hall currents (HCs). This scheme, as visualized in Fig. 1, is identical to the experimental scheme described in the main article. A non-centrosymmetric semiconductor, e.g., GaAs, is excited at normal incidence with circularly-polarized optical pulses of femtosecond duration. The photoexcited carriers will move in the surface electric field, which is perpendicular to the semiconductor surface ($\hat{z}$ direction). An externally applied magnetic field along the $\hat{x}$ direction of the semiconductor leads to Hall rotation of carriers in the $\hat{y}$-$\hat{z}$ plane. This constitutes the HCs. Additionally, due to "anomalous" effects we obtain anomalous current components in all directions, referred to as BE-AHCs. In the following we study the HC along the $\hat{y}$ direction and the BE-AHC along the $\hat{x}$ direction. This is because the two current components can be experimentally separated in free-space THz experiments. (The THz radiation emitted from ultrafast currents shows a polarization in the far field, which is parallel to the direction of the current flow and orthogonal polarizations can be easily separated using THz polarizers.) Thus, this experimental separation of HC and BE-AHC allows a direct comparison between experiment and theory. The THz experiments have two other advantages: they constitute a contactless technique without the need of electrical contacts on the sample and they provide access to currents properties in a broad frequency range.

The remainder of this Supplementary Material is structured as follows. We first comment on the calculation of currents using a density matrix approach and the corresponding velocity matrix elements in Sec. II. We then explain how to set up the Boltzmann equation for carrier density calculations in Sec. III and perform perturbation calculations for electric and magnetic fields in Sec. IV. In Sec. V we establish relations of currents, velocities, and densities for different spin subbands, before we calculate the optically induced carrier distribution in Sec. VI. Expressions for BE-AHCs and HCs are obtained in Sec. VII and Sec. VIII, respectively. Finally, in Sec. IX we calculate the THz responses from valance and conduction bands emitted from BE-AHCs and HCs. These quantities are compared to experiments in the main article.

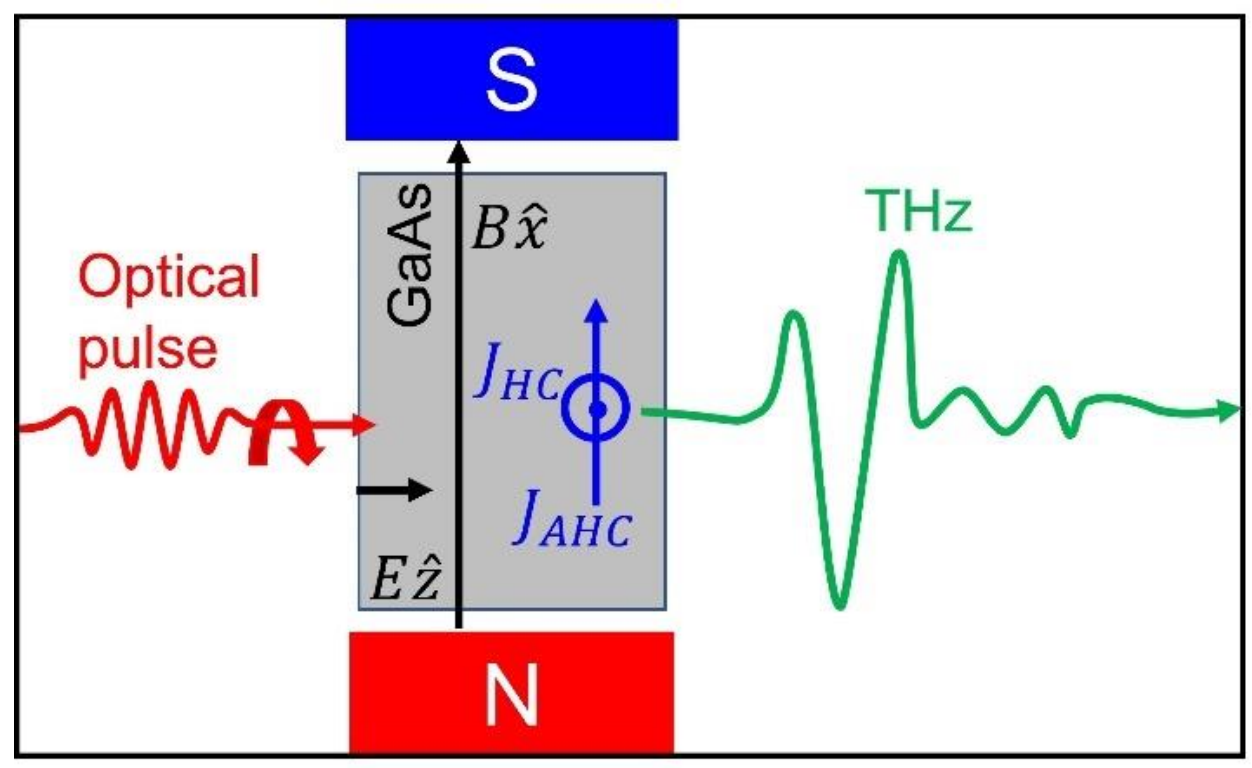

Fig. 1: Scheme for generation of BE-AHC and HC. Circularly-polarized optical pulses excite a GaAs crystal at normal incidence. In the presence of the surface electric field $E$ and an in-plane magnetic field $B$, BE-AHCs ($J_{AHC}$) and HCs ($J_{HC}$) are induced. In this Supplementary Material we focus on $J_{AHC}$ along the $\hat{x}$ direction and $J_{HC}$ along the $\hat{y}$ direction. Since the optical excitation occurs on a femtosecond time scale and the generated current components decay via scattering mechanisms, the currents emit THz radiation, which is used to study the spectral responses of the BE-AHC and the HC.

## II. Calculation of currents and velocities

The generation of BE-AHCs in a bulk GaAs sample involves an optical excitation of electrons and holes with circularly polarized optical pulses. The optically excited carriers then evolve in the sample's surface electric field ($\boldsymbol{E}$) and an externally applied magnetic field ($\boldsymbol{B}$) and lead to BE-AHCs flowing along the direction of $\boldsymbol{B}$. Here, the bulk symbols represent vector quantities. In contrast to the BE-AHCs, HCs flow in the direction of $\boldsymbol{E} \times \boldsymbol{B}$. In the calculation of BE-AHCs and HCs, we ignore any inter-subband coherences, therefore, it is possible to calculate contributions to the currents from an individual band ($\boldsymbol{J}$), only by summing over the corresponding spin-subbands:

$$\boldsymbol{J} = e \sum_{\uparrow,\downarrow} \int d\mathbf{k}\, \boldsymbol{v}_k \rho_{\mathbf{k}} = e \int d\mathbf{k}\, (\boldsymbol{v}_{k,\uparrow} \rho_{\mathbf{k},\uparrow} + \boldsymbol{v}_{-k,\downarrow} \rho_{-\mathbf{k},\downarrow}) \,. \tag{1}$$

Here $e$, $\mathbf{k}$, $\boldsymbol{v}$, and $\rho_k$ are the electron charge, momentum, velocity, and carrier distribution in k-space, respectively, and ↑ and ↓ represent the spin-up and spin-down subbands, respectively. We will use these symbols when we have to distinguish between spin-up and spin-down parameters. For an expression of the velocity $\boldsymbol{v}$, we have to consider that $\boldsymbol{v}$ and $\mathbf{k}$ are related to each other through the following expressions [1]:

$$\boldsymbol{v} = \boldsymbol{v_0} + \boldsymbol{v_{sj}} + \boldsymbol{\Omega} \times \dot{\mathbf{k}} \,, \tag{2}$$

and

$$\dot{\mathbf{k}} = \frac{e}{\hbar}(\boldsymbol{E} + \boldsymbol{v} \times \boldsymbol{B}) \,, \tag{3}$$

where $\boldsymbol{v_0}$, $\boldsymbol{v_{sj}}$, $\boldsymbol{\Omega}$, and $\hbar$, are the normal velocity, side-jump velocity [2–4], Berry curvature, and Planck's constant respectively. The Berry curvature $\boldsymbol{\Omega}$ is an intrinsic property of bandstructures leading to anomalies in carrier transport and is considered to be the main microscopic origin of spin- and anomalous-Hall effects [5]. The side-jump velocity $\boldsymbol{v_{sj}} = \sum_{k'} w_{k'k} \boldsymbol{\Omega} \times (\mathbf{k}' - \mathbf{k})$ is an accumulation of real-space shifts of carriers during scattering processes with $w_{k'k} = w_{kk'}$ being the scattering rate from *k'* to *k* and $\boldsymbol{\Omega} \times (\mathbf{k}' - \mathbf{k})$ being the real-space shift for small scattering angles [6]. Calculating the cross product of the two sides of equation (3) with $\boldsymbol{\Omega}$ yields $\boldsymbol{\Omega} \times \dot{\mathbf{k}} = \frac{e}{\hbar}(\boldsymbol{\Omega} \times \boldsymbol{E} + (\boldsymbol{\Omega} \cdot \boldsymbol{B})\boldsymbol{v} - (\boldsymbol{\Omega} \cdot \boldsymbol{v_0})\boldsymbol{B})$. Inserting this expression into equation (2) yields:

$$\boldsymbol{v} = \varphi_B(\boldsymbol{v_0} + \boldsymbol{v_{sj}} + \frac{e}{\hbar}\boldsymbol{\Omega} \times \boldsymbol{E} - \frac{e}{\hbar}(\boldsymbol{\Omega} \cdot \boldsymbol{v_0})\boldsymbol{B}) \,, \tag{4}$$

with $\varphi_B = 1/(1 - \frac{e}{\hbar}\boldsymbol{\Omega} \cdot \boldsymbol{B})$ .

## III. Setting up the Boltzmann equation for carrier density calculations

The next step will be to calculate the carrier density $\rho_{\mathbf{k}}$. To that end we employ the Boltzmann equation $\dot{\rho_{\mathbf{k}}}(t) + \dot{\rho}_{\mathbf{k},coll.}(t) = -\dot{\mathbf{k}} \cdot \nabla_k \rho_{\mathbf{k}}(t)$. By inserting equation (4) into equation (3), we obtain an expression for $\dot{\mathbf{k}}$:

$$\dot{\mathbf{k}} = \varphi_B \frac{e}{\hbar}(\boldsymbol{E} + (\boldsymbol{v_0} + \boldsymbol{v_{sj}}) \times \boldsymbol{B} - \frac{e}{\hbar}(\boldsymbol{B} \cdot \boldsymbol{E})\boldsymbol{\Omega}) \,. \tag{5}$$

Before we proceed with further calculations, we simplify equations (4) and (5) according to the experimental conditions, i.e., $\boldsymbol{B} = B\hat{x}$ and $\boldsymbol{E} = E\hat{z}$. This yields:

$$\boldsymbol{v} = \varphi_B\left(\boldsymbol{v_0} + \boldsymbol{v_{sj}} + \frac{e}{\hbar}(\hat{x}\Omega^y - \hat{y}\Omega^x)E - \frac{e}{\hbar}(\boldsymbol{\Omega} \cdot \boldsymbol{v_0})B\hat{x}\right) , \tag{6}$$

and

$$\dot{\mathbf{k}} = \frac{e}{\hbar}\varphi_B\left(E\hat{z} + \left(\hat{y}v_0^z - \hat{z}v_0^y\right)B + \left(\hat{y}v_{sj}^z - \hat{z}v_{sj}^y\right)B\right) , \tag{7}$$

with $\varphi_B = \frac{1}{1 - \frac{e}{\hbar}\Omega^x B} \sim (1 + \frac{e}{\hbar}\Omega^x B)$, hereby, ignoring higher orders of $\Omega^x$ in $\varphi_B$.

Inserting equation (7) into the Boltzmann equation results in:

$$\dot{\rho_{\mathbf{k}}}(t) + \dot{\rho}_{\mathbf{k},coll.}(t) = -\frac{e}{\hbar}\varphi_B\left(E\hat{z} + \left(\hat{y}v_0^z - \hat{z}v_0^y\right)B + \left(\hat{y}v_{sj}^z - \hat{z}v_{sj}^y\right)B\right) \cdot \nabla_k \rho_{\mathbf{k}}(t) \,. \tag{8}$$

The collision term for small scattering angles $\theta$ and elastic interactions can be written as $\dot{\rho}_{\mathbf{k},coll.}(t) = \sum_{k'} w_{k'k}(\rho_{\mathbf{k}}(t) - \rho_{\mathbf{k}'}(t)) \sim \nabla_k \rho_{\mathbf{k}}(t) \cdot \sum_{k'} w_{k'k}(\mathbf{k} - \mathbf{k}') + \frac{\mathrm{d}\rho_{\mathbf{k}}(t)}{\mathrm{d}\epsilon_{\mathbf{k}}} \sum_{k'} w_{k'k}(\epsilon_{\mathbf{k}} - \epsilon_{\mathbf{k}'})$ [2,3,7]. The first term on the right-hand side represents carrier scattering processes without loss of energy ($\epsilon_{\mathbf{k}}$) of the involved carriers. The second term on the right-hand side represents carrier scattering processes in which the carriers loose or gain energy in an external electric field. With $\mathbf{k}' = k' \cos(\theta)\,\hat{k} + k' \sin(\theta)\hat{k}_\perp$, and $\hat{k}$ and $\hat{k}_\perp$ being unit vectors along and perpendicular, respectively, to $\mathbf{k}$, the collision term can be further simplified to:

$$\dot{\rho}_{\mathbf{k},coll.}(t) = \nabla_{\mathbf{k}}\rho_{\mathbf{k}}(t) \cdot \sum_{k'} w_{k'k}\left\{(k - k'\cos(\theta))\hat{k} - k'\sin(\theta)\hat{k}_{\perp}\right\} - {}^{e}/_{\hbar}\,(\boldsymbol{\nabla}_{\mathbf{k}}\rho_{\mathbf{k}}(t) \cdot {}^{1}/_{\boldsymbol{v}_{\mathbf{0}}})\boldsymbol{E} \cdot \sum_{\mathbf{k}'} w_{k'k}\ \boldsymbol{\Omega} \times (\mathbf{k}' - \mathbf{k})$$

$$= \gamma\rho_{\mathbf{k}}(t) - w^{a}Bk\nabla_{\mathbf{k}}\rho_{\mathbf{k}}(t) \cdot \hat{\boldsymbol{k}}_{\perp} - {}^{e}/_{\hbar}\,(\boldsymbol{\nabla}_{\mathbf{k}}\rho_{\mathbf{k}}(t) \cdot {}^{1}/_{\boldsymbol{v}_{\mathbf{0}}})Ev_{sj}^{z}\ . \tag{9}$$

Here, $\gamma = \sum_{k'} w_{k'k}(1 - \cos(\theta))$ is the symmetric momentum scattering rate. However, because of the magnetic field, $w_{k'k}$ becomes an asymmetric function of $k'$ [8]. Such an asymmetry results in the survival of the $\sin(\theta)\hat{k}_{\perp}$term along the direction of $B$ and leads to skew scattering of carriers along this direction. With these definitions we obtain $w^{a} = \frac{1}{B}\sum_{k'} w_{k'k}\sin(\theta)\hat{k}_{\perp}$ as the coefficient of asymmetric scattering being first order in $B$. We consider any $B$ independent skew scattering to be very weak and consequently ignore it [3]. The last term in equation (9) is called the anomalous distribution term [4,6]. In this term, we ignore any asymmetry in $w_{k'k}$ and utilize the fact that the change in energy equals the work done by $E$ during the real-space shift. With the above assumptions, we can rewrite equation (8):

$$\dot{\rho}_{\mathbf{k}}(t) + \gamma\rho_{\mathbf{k}}(t) = -\frac{e}{\hbar}\Big(\varphi_{B}E\partial_{z}\rho_{\mathbf{k}}(t) + \left(v_{0}^{z}\partial_{y}\rho_{\mathbf{k}}(t) - v_{0}^{y}\partial_{z}\rho_{\mathbf{k}}(t)\right)\varphi_{B}B + \left(\hat{y}v_{sj}^{z} - \hat{z}v_{sj}^{y}\right)\varphi_{B}B - Ev_{sj}^{z}\nabla_{\mathbf{k}}\rho_{\mathbf{k}}(t) \cdot {}^{1}/_{\boldsymbol{v}_{\mathbf{0}}}\Big) + w_{a}Bk\partial_{x}\rho_{\mathbf{k}}(t)\ , \tag{10}$$

where $\partial_{x,y,z} = \frac{\partial}{\partial k^{x},\partial k^{y},\partial k^{z}}$ .

## IV. Perturbation solution of carrier densities in external fields

We solve equation (10) employing perturbation solutions for $B$ and $E$ for one spin subband and set the optically excited spin-polarized carriers as zero-order solution $\rho_{\mathbf{k}}^{(0)}$. Since in the experiment, we observe spectral features occurring at the cyclotron frequency $\omega_{B} = \frac{e}{m}B$ with $m$ being the effective mass of the carriers, we first determine the carrier population rotating in the magnetic field. To that end, we calculate the perturbation solution of equation (10) up to infinite order in $B$ and for $0^{\text{th}}$ order in $E$, i.e., we set $E = 0$ in equation (10). Without solving equation (10) up to infinite order in $B$, carrier rotation in $B$ cannot be accounted for. It is important to note that in the following calculations, we will only consider anomalous coefficients, i.e., $\boldsymbol{\Omega}$, $\boldsymbol{v_{sj}}$, and $w_{a}$, up to their first order. We will ignore momentum-space derivatives of anomalous coefficients and multiplication of anomalous coefficients with each other. With these assumptions, equation (10) reduces to:

$$\dot{\rho}_{\mathbf{k}}^{(B)}(t) + \gamma\rho_{\mathbf{k}}^{(B)}(t) = \frac{m}{\hbar}\omega_{B}\varphi_{B}\left(v_{0}^{y}\partial_{z}\rho_{\mathbf{k}}^{(B)}(t) - v_{0}^{z}\partial_{y}\rho_{\mathbf{k}}^{(B)}(t)\right) + \frac{m}{\hbar}\omega_{B}\left(v_{sj}^{y}\partial_{z}\rho_{\mathbf{k}}^{(B)}(t) - v_{sj}^{z}\partial_{y}\rho_{\mathbf{k}}^{(B)}(t)\right) + w_{a}Bk\partial_{x}\rho_{\mathbf{k}}^{(B)}(t)\ . \tag{11}$$

In the following, we keep the skew scattering only to the first order in $B$, ignore any derivative of $\partial_{x,y,z}\rho_{\mathbf{k}}^{(0)}$ and $\Omega^{x}$, and assume $\partial_{x,y,z}v_{0}^{x,y,z} = \frac{\hbar}{m}$ and $\partial_{x}v_{0}^{y,z} = 0$. Employing Fourier transformation, equation (11) can be written in the frequency domain as:

$$\rho_{\mathbf{k}}^{(B)} = \frac{m}{\hbar}\frac{\omega_{B}}{s}\left(v_{0}^{y}\partial_{z}\rho_{\mathbf{k}}^{(B)} - v_{0}^{z}\partial_{y}\rho_{\mathbf{k}}^{(B)}\right)\left(1 + \frac{e}{\hbar}\Omega^{x}B\right), \tag{12}$$

where $s = i\omega + \gamma$. In equation (12), the last two terms of equation (11), i.e., the side-jump and the skew-scattering terms, have been omitted. We will add the skew scattering term $\frac{w_a B}{s} k \partial_x \rho_{\mathbf{k}}^{(0)}$ to $\rho_{\mathbf{k}}^{(B)}$ after the perturbation calculation, since we consider skew scattering terms only up to the first order in $B$. We will also add the side-jump term to $\rho_{\mathbf{k}}^{(B)}$ after the perturbation process, since it survives only up to the first order in $B$. To simplify the necessary steps required to obtain $\rho_{\mathbf{k}}^{(B)}$ from equation (12), we define $D = \frac{m}{\hbar}\left(v_0^y \partial_z - v_0^z \partial_y\right)$, $X = D\rho_{\mathbf{k}}^{(0)} = \frac{m}{\hbar}\left(v_0^y \partial_z \rho_{\mathbf{k}}^{(0)} - v_0^z \partial_y \rho_{\mathbf{k}}^{(0)}\right)$, $Y = -DX = \frac{m}{\hbar}\left(v_0^y \partial_y \rho_{\mathbf{k}}^{(0)} + v_0^z \partial_z \rho_{\mathbf{k}}^{(0)}\right)$, $a = \frac{\omega_B}{s}$, and $\alpha = \frac{e}{\hbar}\Omega^x B$. Using the introduced terminology, it is easy to obtain $X = DY$ and we rewrite equation (12) as:

$$\rho^{(B)} = a(1+\alpha)D\rho^{(B)}\,, \tag{13}$$

Employing perturbation techniques, we obtain $\rho^{(B^1)} = aX$, $\rho^{(B^2)} = -a^2 Y + a\alpha X$, $\rho^{(B^3)} = -a^3 X - 2a^2\alpha Y$, $\rho^{(B^4)} = a^4 Y - 3a^3\alpha X$, $\rho^{(B^5)} = a^5 X + 4a^4 \alpha Y$, $\rho^{(B^6)} = -a^6 Y + 5a^5 \alpha X$, and so on. Calculating till $\rho^{(B^\infty)}$, we obtain:

$$\begin{aligned}
\rho^{(B)} &= \rho^{(B^1)} + \rho^{(B^2)} + \cdots + \rho^{(B^\infty)} \\
&= aX(1 - a^2 + a^4 - \cdots) - a^2 Y(1 - a^2 + a^4 - \cdots) + a\alpha X(1 - 3a^2 + 5a^4 + \cdots) \\
&\qquad - 2a^2\alpha Y(1 - 2a^2 + 3a^4 - \cdots) \\
&= \frac{aX}{1+a^2} - \frac{a^2 Y}{1+a^2} + \alpha\frac{a(1-a^2)X}{(1+a^2)^2} - \alpha\frac{2a^2 Y}{(1+a^2)^2} \\
&= \frac{aX}{1+a^2}\left(1 + \alpha\frac{(1-a^2)}{1+a^2}\right) - \frac{a^2 Y}{1+a^2}\left(1 + \alpha\frac{2}{1+a^2}\right).
\end{aligned}$$

Now we rewrite the expression for $\rho^{(B)}$ using the initially defined terms:

$$\begin{aligned}
\rho_{\mathbf{k}}^{(B)} &= \frac{m}{\hbar}\left(v_0^y \partial_z \rho_{\mathbf{k}}^{(0)} - v_0^z \partial_y \rho_{\mathbf{k}}^{(0)}\right)\frac{s\omega_B}{(s^2+\omega_B^2)}\left(1 + \frac{e}{\hbar}\Omega^x B\frac{(s^2-\omega_B^2)}{(s^2+\omega_B^2)}\right) - \frac{m}{\hbar}\left(v_0^z \partial_z \rho_{\mathbf{k}}^{(0)} + \right. \\
&\left. v_0^y \partial_y \rho_{\mathbf{k}}^{(0)}\right)\frac{\omega_B^2}{(s^2+\omega_B^2)}\left(1 + \frac{e}{\hbar}\Omega^x B\frac{2s^2}{(s^2+\omega_B^2)}\right) + \frac{w_a B}{s} k \partial_x \rho_{\mathbf{k}}^{(0)} + \frac{m}{\hbar}\frac{\omega_B}{s}\left(v_{sj}^y \partial_z \rho_{\mathbf{k}}^{(0)}(t) - \right. \\
&\left. v_{sj}^z \partial_y \rho_{\mathbf{k}}^{(0)}(t)\right).
\end{aligned} \tag{14a}$$

This equation constitutes the perturbation solution of equation (11) up to infinite order in $B$ and includes the side-jump and skew-scattering terms which have been omitted during the perturbation calculation. We note that equation (14a) still constitutes a frequency expression.

As a next step, we calculate the population being first order in $E$ and infinite order in $B$ by rewriting equation (10) as follows:

$$\dot{\rho}_{\mathbf{k}}^{(E^1,B)}(t) + \gamma\rho_{\mathbf{k}}^{(E^1,B)}(t) = -\frac{e}{\hbar}\left(\left(1 + \frac{e}{\hbar}\Omega^x B\right)E\partial_z \rho_{\mathbf{k}}^{(B)}(t) - E v_{sj}^z \nabla_{\mathbf{k}} \rho_{\mathbf{k}}^{(B)}(t)\cdot {}^{1}\!/_{\boldsymbol{v}_0}\right).$$

Here, on the right-hand side of the equation, $\rho_{\mathbf{k}}(t)$ has been replaced by $\rho_{\mathbf{k}}^{(B)}(t)$ and we only consider terms depending on $E$. After a Fourier transformation of the last equation and expressing $\rho_{\mathbf{k}}^{(B)}$ with equation (14a) we obtain in the frequency domain:

$$\rho_{\mathbf{k}}^{(E^1,B)} = \frac{e}{\hbar}E\left(\frac{\omega_B}{(s^2+\omega_B^2)}\left(1+\frac{e}{\hbar}\Omega^x B\,\frac{2s^2}{(s^2+\omega_B^2)}\right)\partial_y\rho_{\mathbf{k}}^{(0)} + \frac{\omega_B^2}{s(s^2+\omega_B^2)}\left(1+\frac{e}{\hbar}\Omega^x B\,\frac{3s^2+\omega_B^2}{(s^2+\omega_B^2)}\right)\partial_z\rho_{\mathbf{k}}^{(0)}\right) + \frac{e}{\hbar}Ev_{sj}^z\left\{\left(\frac{\partial_z\rho_{\mathbf{k}}^{(0)}}{v_0^y} - \frac{\partial_y\rho_{\mathbf{k}}^{(0)}}{v_0^z}\right)\frac{\omega_B}{(s^2+\omega_B^2)} - \left(\frac{\partial_y\rho_{\mathbf{k}}^{(0)}}{v_0^y} + \frac{\partial_z\rho_{\mathbf{k}}^{(0)}}{v_0^z}\right)\frac{\omega_B^2}{s(s^2+\omega_B^2)}\right\} - \frac{e}{\hbar}E\,\frac{k^z}{k}\,\partial_x\rho_{\mathbf{k}}^{(0)}\,\frac{w_a B}{s^2}\,. \tag{14b}$$

In addition to $\rho_{\mathbf{k}}^{(B)}$ and $\rho_{\mathbf{k}}^{(E^1,B)}$, we also require an expression for the population which is first order in $E$ and zeroth order in $B$, i.e., $\rho_{\mathbf{k}}^{(E^1)}$, which is obtained by solving:

$$\dot{\rho}_{\mathbf{k}}^{(E^1)}(t) + \gamma\rho_{\mathbf{k}}^{(E^1)}(t) = -\frac{e}{\hbar}\left(E\partial_z\rho_{\mathbf{k}}^{(0)}(t) - Ev_{sj}^z\nabla_{\mathbf{k}}\rho_{\mathbf{k}}^{(0)}(t)\cdot {}^{1}/_{\boldsymbol{v_0}}\right),$$

As frequency-domain expression for $\rho_{\mathbf{k}}^{(E^1)}$ we obtain:

$$\rho_{\mathbf{k}}^{(E^1)} = -\frac{e}{\hbar}E\partial_z\rho_{\mathbf{k}}^{(0)}\frac{1}{s} + \frac{e}{\hbar}Ev_{sj}^z\left(\frac{\partial_x\rho_{\mathbf{k}}^{(0)}}{v_0^x} + \frac{\partial_y\rho_{\mathbf{k}}^{(0)}}{v_0^y} + \frac{\partial_z\rho_{\mathbf{k}}^{(0)}}{v_0^z}\right)\frac{1}{s}\,. \tag{14c}$$

## V. Currents, velocities, and densities for different spin subbands

As introduced in equation (1), we consider both spin subbands in the current calculations. From equation (7), we obtain the spin-up and spin-down velocities along x-direction as:

$$v_{k,\uparrow}^x = v_{sj,k,\uparrow}^x + v_{0,k,\uparrow}^x + \frac{e}{\hbar}\Omega_{k,\uparrow}^y E - \frac{e}{\hbar}\left(\Omega_{k,\uparrow}^y v_{0,k,\uparrow}^y + \Omega_{k,\uparrow}^z v_{0,k,\uparrow}^z\right)B\,, \tag{15a}$$

$$v_{-k,\downarrow}^x = v_{sj,k,\uparrow}^x - v_{0,k,\uparrow}^x - \frac{e}{\hbar}\Omega_{k,\uparrow}^y E - \frac{e}{\hbar}\left(\Omega_{k,\uparrow}^y v_{0,k,\uparrow}^y + \Omega_{k,\uparrow}^z v_{0,k,\uparrow}^z\right)B\,. \tag{15b}$$

Here the two velocities are linked via the following symmetry relations: $v_{0,k,\uparrow} = -v_{0,-k,\downarrow}$, $\Omega_{k,\uparrow} = -\Omega_{-k,\downarrow}$, and $v_{sj,k,\uparrow} = v_{sj,-k,\downarrow}$ [3],[9]. Inserting equations (15a) and (15b) into equation (1) yields:

$$J^x = e\int d\mathbf{k}\left\{\left(v_{sj,k,\uparrow}^x - \frac{e}{\hbar}\left(\Omega_{k,\uparrow}^y v_{0,k,\uparrow}^y + \Omega_{k,\uparrow}^z v_{0,k,\uparrow}^z\right)B\right)\left(\rho_{\mathbf{k},\uparrow} + \rho_{-\mathbf{k},\downarrow}\right) + \left(v_{0,k,\uparrow}^x + \frac{e}{\hbar}\Omega_{k,\uparrow}^y E\right)\left(\rho_{\mathbf{k},\uparrow} - \rho_{-\mathbf{k},\downarrow}\right)\right\}.$$

Since we are only interested in currents depending on $E$ and $B$, we rewrite $J^x$ with terms having both $E$ and $B$, thereby considering the densities $\rho_{\mathbf{k}}^{(B)}$, $\rho_{\mathbf{k}}^{(E^1,B)}$, and $\rho_{\mathbf{k}}^{(E^1)}$:

$$J^x = e\int d\mathbf{k}\left\{v_{sj,k,\uparrow}^x\left(\rho_{\mathbf{k},\uparrow}^{(E^1,B)} + \rho_{-\mathbf{k},\downarrow}^{(E^1,B)}\right) - \frac{e}{\hbar}\left(\Omega_{k,\uparrow}^y v_{0,k,\uparrow}^y + \Omega_{k,\uparrow}^z v_{0,k,\uparrow}^z\right)B\left(\rho_{\mathbf{k},\uparrow}^{(E^1,B)} + \rho_{\mathbf{k},\uparrow}^{(E^1)} + \rho_{-\mathbf{k},\downarrow}^{(E^1,B)} + \rho_{-\mathbf{k},\downarrow}^{(E^1)}\right) + v_{0,k,\uparrow}^x\left(\rho_{\mathbf{k},\uparrow}^{(E^1,B)} - \rho_{-\mathbf{k},\downarrow}^{(E^1,B)}\right) + \frac{e}{\hbar}\Omega_{k,\uparrow}^y E\left(\rho_{\mathbf{k},\uparrow}^{(B)} - \rho_{-\mathbf{k},\downarrow}^{(B)}\right)\right\}. \tag{16a}$$

Here, we emphasize that the currents are only first order in $E$. Equation (16a) expresses BE-AHCs, which flow along the x-direction for the external fields $E$ and $B$ being parallel to z and x, respectively. A similar expression for HCs, which flow along the y-direction can be obtained from:

$$J^y = e\int d\boldsymbol{k}\,v_{0,k,\uparrow}^y\left(\rho_{\mathbf{k},\uparrow}^{(E^1,B)} - \rho_{-\mathbf{k},\downarrow}^{(E^1,B)}\right). \tag{16b}$$

For HCs we ignore Berry-curvature and side-jump contributions to $\boldsymbol{v}$.

To obtain more complete expressions for the BE-AHC using equation (16a) and for the HC using equation (16b) we need to analyze the terms $\rho_{\mathbf{k},\uparrow}^{(B)} - \rho_{-\mathbf{k},\downarrow}^{(B)}$, $\rho_{\mathbf{k},\uparrow}^{(E^1,B)} - \rho_{-\mathbf{k},\downarrow}^{(E^1,B)}$, $\rho_{\mathbf{k},\uparrow}^{(E^1,B)} + \rho_{-\mathbf{k},\downarrow}^{(E^1,B)}$, and $\rho_{\mathbf{k},\uparrow}^{(E^1)} + \rho_{-\mathbf{k},\downarrow}^{(E^1)}$ using spin-up and spin-down versions of equations (14). Using the symmetry relation $w^a = w_{k,\uparrow}^a = w_{-k,\downarrow}^a$ and the above-mentioned symmetry relations it follows from equation (14a):

$$\rho_{\mathbf{k},\uparrow}^{(B)} - \rho_{-\mathbf{k},\downarrow}^{(B)} = \frac{e}{\hbar}\frac{B}{(s^2+\omega_B^2)}\left\{s\left(v_{0,k,\uparrow}^{y}\partial_z - v_{0,k,\uparrow}^{z}\partial_y\right) - \omega_B\left(v_{0,k,\uparrow}^{z}\partial_z + v_{0,k,\uparrow}^{y}\partial_y\right)\right\}\left(\rho_{\mathbf{k},\uparrow}^{(0)} - \rho_{-\mathbf{k},\downarrow}^{(0)}\right) + \left\{\frac{e^2}{\hbar^2}\frac{sB^2\Omega_{k,\uparrow}^{x}}{(s^2+\omega_B^2)^2}\left\{(s^2-\omega_B^2)\left(v_{0,k,\uparrow}^{y}\partial_z - v_{0,k,\uparrow}^{z}\partial_y\right) - 2s\omega_B\left(v_{0,k,\uparrow}^{z}\partial_z + v_{0,k,\uparrow}^{y}\partial_y\right)\right\} + \frac{w_a B}{s}k\partial_x\right\}\left(\rho_{\mathbf{k},\uparrow}^{(0)} + \rho_{-\mathbf{k},\downarrow}^{(0)}\right) + \frac{m}{\hbar}\frac{\omega_B}{s}\left(v_{sj,\uparrow}^{y}\partial_z - v_{sj,\uparrow}^{z}\partial_y\right)\left(\rho_{\mathbf{k},\uparrow}^{(0)} + \rho_{-\mathbf{k},\downarrow}^{(0)}\right). \tag{17a}$$

Likewise, with the discussed symmetry relations and equation (14b), we obtain:

$$\rho_{\mathbf{k},\uparrow}^{(E^1,B)} - \rho_{-\mathbf{k},\downarrow}^{(E^1,B)} = \frac{e}{\hbar}E\left\{\frac{e}{\hbar}\frac{\omega_B\Omega_{k,\uparrow}^{x}B}{(s^2+\omega_B^2)^2}\left(2s^2\partial_y + \frac{\omega_B(3s^2+\omega_B^2)}{s}\partial_z\right) + \frac{\omega_B v_{sj,k,\uparrow}^{z}}{(s^2+\omega_B^2)}\left(\left(\frac{\partial_z}{v_{0,k,\uparrow}^{y}} - \frac{\partial_y}{v_{0,k,\uparrow}^{z}}\right) - \frac{\omega_B}{s}\left(\frac{\partial_y}{v_{0,k,\uparrow}^{y}} + \frac{\partial_z}{v_{0,k,\uparrow}^{z}}\right)\right) - \frac{k^z}{k}\frac{w_a B}{s^2}\partial_x\right\}\left(\rho_{\mathbf{k},\uparrow}^{(0)} - \rho_{-\mathbf{k},\downarrow}^{(0)}\right) + \frac{e}{\hbar}\frac{\omega_B E}{(s^2+\omega_B^2)}\left\{\partial_y + \frac{\omega_B}{s}\partial_z\right\}\left(\rho_{\mathbf{k},\uparrow}^{(0)} + \rho_{-\mathbf{k},\downarrow}^{(0)}\right). \tag{17b}$$

and

$$\rho_{\mathbf{k},\uparrow}^{(E^1,B)} + \rho_{-\mathbf{k},\downarrow}^{(E^1,B)} = \frac{e}{\hbar}E\left\{\frac{e}{\hbar}\frac{\omega_B\Omega_{k,\uparrow}^{x}B}{(s^2+\omega_B^2)^2}\left(2s^2\partial_y + \frac{\omega_B(3s^2+\omega_B^2)}{s}\partial_z\right) + \frac{\omega_B v_{sj,k,\uparrow}^{z}}{(s^2+\omega_B^2)}\left(\left(\frac{\partial_z}{v_{0,k,\uparrow}^{y}} - \frac{\partial_y}{v_{0,k,\uparrow}^{z}}\right) - \frac{\omega_B}{s}\left(\frac{\partial_y}{v_{0,k,\uparrow}^{y}} + \frac{\partial_z}{v_{0,k,\uparrow}^{z}}\right)\right) - \frac{k^z}{k}\frac{w_a B}{s^2}\partial_x\right\}\left(\rho_{\mathbf{k},\uparrow}^{(0)} + \rho_{-\mathbf{k},\downarrow}^{(0)}\right) + \frac{e}{\hbar}\frac{\omega_B E}{(s^2+\omega_B^2)}\left\{\partial_y + \frac{\omega_B}{s}\partial_z\right\}\left(\rho_{\mathbf{k},\uparrow}^{(0)} - \rho_{-\mathbf{k},\downarrow}^{(0)}\right). \tag{17c}$$

Finally, with the discussed symmetry relations and equation (14c), we obtain:

$$\rho_{\mathbf{k},\uparrow}^{(E^1)} + \rho_{-\mathbf{k},\downarrow}^{(E^1)} = \frac{e}{\hbar}\frac{1}{s}E v_{sj,k,\uparrow}^{z}\left(\frac{\partial_x}{v_{0,k,\uparrow}^{x}} + \frac{\partial_y}{v_{0,k,\uparrow}^{y}} + \frac{\partial_z}{v_{0,k,\uparrow}^{z}}\right)\left(\rho_{\mathbf{k},\uparrow}^{(0)} + \rho_{-\mathbf{k},\downarrow}^{(0)}\right) - \frac{e}{\hbar}\frac{1}{s}E\partial_z\left(\rho_{\mathbf{k},\uparrow}^{(0)} - \rho_{-\mathbf{k},\downarrow}^{(0)}\right). \tag{17d}$$

## VI. Optically excited carriers

Before we proceed with the calculations of BE-AHCs and HCs, we examine the terms $(\rho_{\mathbf{k},\uparrow}^{(0)} - \rho_{-\mathbf{k},\downarrow}^{(0)})$ and $(\rho_{\mathbf{k},\uparrow}^{(0)} + \rho_{-\mathbf{k},\downarrow}^{(0)})$ appearing in equations (17). In the experiments both, BE-AHC and HC, are induced upon circularly polarized optical excitation. In semiconductors, two types of carrier distributions appear upon excitation with circularly polarized light:

(a) A carrier distribution sensitive to the helicity of the circular polarization of the light, which results in BE-AHC. In the following this distribution is referred to as type-a distribution. The initial k-space distribution (distribution without any scattering) for different spins of type-a carriers relate to each other as $g_{c,\mathbf{k},\uparrow} = -g_{c,-\mathbf{k},\downarrow}$.

(b) A carrier distribution resulting from orthogonal polarization components of the circular polarization and being insensitive of the helicity of the circular polarization of the light [9]. In the following this distribution is referred to as type-b distribution. The type-b carriers are the main contributors to the HC, since it does not depend on the helicity of the optical excitation as observed experimentally. (We note that a small BE-AHC contribution will occur in the direction of HC, which we could not experimentally extract.) The initial k-space distributions for different spins of the type-b carriers relate to each other as $g_{l,\mathbf{k},\uparrow} = g_{l,-\mathbf{k},\downarrow}$.

In the following we will derive expressions for $(\rho_{\mathbf{k},\uparrow}^{(0)} - \rho_{-\mathbf{k},\downarrow}^{(0)})$ and $(\rho_{\mathbf{k},\uparrow}^{(0)} + \rho_{-\mathbf{k},\downarrow}^{(0)})$, which are needed for the calculation of BE-AHC and HC, respectively. The term $\rho_{\mathbf{k},\uparrow}^{(0)}$ is obtained by solving the density-matrix equation [9]:

$$\dot{\rho}_{\mathbf{k},\uparrow}^{(0)}(t) = g_{\mathbf{k},\uparrow} I(t) - \dot{\rho}_{\mathbf{k},\uparrow,coll.}^{(0)}(t) \,. \tag{18}$$

Here, $g_{\mathbf{k},\uparrow} = g_{c,\mathbf{k},\uparrow} + g_{l,\mathbf{k},\uparrow}$ is the initial spin-up carrier distribution in k-space considering both, type-a and type-b carriers and $I(t)$ is the optical pulse intensity. To solve equation (18) we adopt the relaxation time approximation for the collision term $\dot{\rho}_{\mathbf{k},\uparrow,coll.}^{(0)}(t) = \gamma\left(\rho_{\mathbf{k},\uparrow}^{(0)}(t) - \bar{\rho}_{\mathbf{k},\uparrow}^{(0)}(t)\right) + \frac{w^{\mathrm{sf}}}{2}(\bar{\rho}_{\mathbf{k},\uparrow}^{(0)}(t) - \bar{\rho}_{-\mathbf{k},\downarrow}^{(0)}(t))$, with $\bar{\rho}_{\mathbf{k},\uparrow}^{(0)}(t)$ denoting the momentum-relaxed population, which decays only with the spin-flip relaxation rate $w^{\mathrm{sf}}$. We rewrite equation (18) for spin-up and -down subbands as:

$$\dot{\rho}_{\mathbf{k},\uparrow}^{(0)}(t) = g_{\mathbf{k},\uparrow} I(t) - \gamma\left(\rho_{\mathbf{k},\uparrow}^{(0)}(t) - \bar{\rho}_{\mathbf{k},\uparrow}^{(0)}(t)\right) - \frac{w^{\mathrm{sf}}}{2}(\bar{\rho}_{\mathbf{k},\uparrow}^{(0)}(t) - \bar{\rho}_{-\mathbf{k},\downarrow}^{(0)}(t)) \,, \tag{19a}$$

$$\dot{\rho}_{-\mathbf{k},\downarrow}^{(0)}(t) = g_{-\mathbf{k},\downarrow} I(t) - \gamma\left(\rho_{-\mathbf{k},\downarrow}^{(0)}(t) - \bar{\rho}_{-\mathbf{k},\downarrow}^{(0)}(t)\right) + \frac{w^{\mathrm{sf}}}{2}(\bar{\rho}_{\mathbf{k},\uparrow}^{(0)}(t) - \bar{\rho}_{-\mathbf{k},\downarrow}^{(0)}(t)) \,. \tag{19b}$$

Now we first take the average on both sides of equations (19a) and (19b), and then subtract the two equations from each other to obtain:

$$\dot{\bar{\rho}}_{\mathbf{k},\uparrow}^{(0)}(t) - \dot{\bar{\rho}}_{-\mathbf{k},\downarrow}^{(0)}(t) = (\bar{g}_{\mathbf{k},\uparrow} - \bar{g}_{-\mathbf{k},\downarrow}) I(t) - w^{\mathrm{sf}}\left(\bar{\rho}_{\mathbf{k},\uparrow}^{(0)}(t) - \bar{\rho}_{-\mathbf{k},\downarrow}^{(0)}(t)\right) .$$

Using Fourier transformation the differential equation can to solved, yielding:

$$\bar{\rho}_{\mathbf{k},\uparrow}^{(0)}(t) - \bar{\rho}_{-\mathbf{k},\downarrow}^{(0)}(t) = (\bar{g}_{\mathbf{k},\uparrow} - \bar{g}_{-\mathbf{k},\downarrow}) I(t) * u(t)\exp\left(-w^{\mathrm{sf}} t\right) \tag{20}$$

Directly subtracting equations (19a) and (19b) from each other (without averaging), we obtain:

$$\dot{\rho}_{\mathbf{k},\uparrow}^{(0)}(t) - \dot{\rho}_{-\mathbf{k},\downarrow}^{(0)}(t) = (g_{\mathbf{k},\uparrow} - g_{-\mathbf{k},\downarrow}) I(t) - \gamma\left(\rho_{\mathbf{k},\uparrow}^{(0)}(t) - \rho_{-\mathbf{k},\downarrow}^{(0)}(t)\right) + \left(\gamma - w^{\mathrm{sf}}\right)\left(\bar{\rho}_{\mathbf{k},\uparrow}^{(0)}(t) - \bar{\rho}_{-\mathbf{k},\downarrow}^{(0)}(t)\right) .$$

Again, we can solve the differential equation using Fourier transformation. Using equation (20) we obtain:

$$\rho_{\mathbf{k},\uparrow}^{(0)}(t) - \rho_{-\mathbf{k},\downarrow}^{(0)}(t) = (g_{\mathbf{k},\uparrow} - g_{-\mathbf{k},\downarrow}) I(t) * u(t)\exp(-\gamma t) + \left(\gamma - w^{\mathrm{sf}}\right)(\bar{g}_{\mathbf{k},\uparrow} - \bar{g}_{-\mathbf{k},\downarrow}) I(t) * u(t)\exp\left(-w^{\mathrm{sf}} t\right) * u(t)\exp(-\gamma t)$$

Expressing $g_{\mathbf{k}}$ with $g_{c,\mathbf{k}}$ and $g_{l,\mathbf{k}}$ and considering the symmetry relations, we obtain in the time domain:

$\rho^{(0)}_{\mathbf{k},\uparrow}(t) - \rho^{(0)}_{-\mathbf{k},\downarrow}(t) = 2\big(g_{c,\mathbf{k},\uparrow} - \bar{g}_{c,\mathbf{k},\uparrow}\big) I(t) * u(t)\exp(-\gamma t) + 2\bar{g}_{c,\mathbf{k},\uparrow} I(t) *$
$u(t)\exp\big(-w^{\mathrm{sf}} t\big),$

and in the frequency domain:

$$\rho^{(0)}_{\mathbf{k},\uparrow} - \rho^{(0)}_{-\mathbf{k},\downarrow} = 2\big(g_{c,\mathbf{k},\uparrow} - \bar{g}_{c,\mathbf{k},\uparrow}\big)\frac{I(\omega)}{s} + 2\bar{g}_{c,\mathbf{k},\uparrow}\frac{I(\omega)}{i\omega + w^{\mathrm{sf}}}\,.$$

In the above equation we have assumed that the term $\bar{g}_{l,\mathbf{k},\uparrow}$ follows a symmetry relation similar to $g_{l,\mathbf{k},\uparrow}$, i.e., $\bar{g}_{l,\mathbf{k},\uparrow} = \bar{g}_{l,-\mathbf{k},\downarrow}$. The term $\big(g_{c,\mathbf{k},\uparrow} - \bar{g}_{c,\mathbf{k},\uparrow}\big)$ accounts for a k-space asymmetry in optically excited carriers, which can be omitted in a simple Zincblende bandstructure with parabolic bands and perfect spin degeneracy. Therefore, we can simplify the last equation:

$$\rho^{(0)}_{\mathbf{k},\uparrow} - \rho^{(0)}_{-\mathbf{k},\downarrow} = 2\bar{g}_{c,\mathbf{k},\uparrow}\frac{I(\omega)}{i\omega + w^{\mathrm{sf}}}\,. \tag{21}$$

Equation (21) will be used for the calculation of BE-AHCs, since it accounts for the aforementioned type-a carriers.

To calculate the behavior of type-b carriers, we add equations (19a) and (19b) and obtain:

$$\dot{\rho}^{(0)}_{\mathbf{k},\uparrow}(t) + \dot{\rho}^{(0)}_{-\mathbf{k},\downarrow}(t) = 2g_{l,\mathbf{k},\uparrow} I(t) - w^{\mathrm{E}}\left(\rho^{(0)}_{\mathbf{k},\uparrow}(t) - \bar{\rho}^{(0)}_{\mathbf{k},\uparrow}(t) + \rho^{(0)}_{-\mathbf{k},\downarrow}(t) - \bar{\rho}^{(0)}_{-\mathbf{k},\downarrow}(t)\right). \tag{22}$$

In equation (22) we have replaced $\gamma$ with an energy relaxation rate $w^{\mathrm{E}}$. The reason for choosing $w^{\mathrm{E}}$ as a relaxation mechanism for the HC generation process, is the assumption that HCs are strong for hot carriers ($g_{l,\mathbf{k},\uparrow}$) and significantly decrease for cold carriers ($\bar{g}_{l,\mathbf{k},\uparrow}$). We take the average on the two sides of equation (22) and solve using Fourier transformation to obtain:

$$\bar{\rho}^{(0)}_{\mathbf{k},\uparrow}(t) + \bar{\rho}^{(0)}_{-\mathbf{k},\downarrow}(t) = 2\bar{g}_{l,\mathbf{k},\uparrow} I(t) * u(t)\,. \tag{23}$$

Here we have considered the recombination time to be infinite as it is much longer than all other scattering times and, consequently does not influence the dynamics on ultrafast time scales. Inserting equation (23) in equation (22), we obtain in the time domain:

$$\rho^{(0)}_{\mathbf{k},\uparrow}(t) + \rho^{(0)}_{-\mathbf{k},\downarrow}(t) = 2(g_{l,\mathbf{k},\uparrow} - \bar{g}_{l,\mathbf{k},\uparrow}) I(t) * u(t)\exp(-w^{\mathrm{E}} t) + 2\bar{g}_{l,\mathbf{k},\uparrow} I(t) * u(t)\,,$$

And in the frequency domain:

$$\rho^{(0)}_{\mathbf{k},\uparrow} + \rho^{(0)}_{-\mathbf{k},\downarrow} = 2(g_{l,\mathbf{k},\uparrow} - \bar{g}_{l,\mathbf{k},\uparrow})\frac{I(\omega)}{i\omega + w^{\mathrm{E}}} + 2\bar{g}_{l,\mathbf{k},\uparrow}\frac{I(\omega)}{i\omega}\,. \tag{24}$$

Equation (24) will be used for the calculation of HCs, since it accounts for the aforementioned type-b carriers.

## VII. Expressions for BE-AHC

In this section we will obtain the final expressions for the BE-AHC. Inserting equation (21) into equations (17) and, consecutively, equations (17) into equation (16a) with omission of the $\rho^{(0)}_{\mathbf{k},\uparrow} + \rho^{(0)}_{-\mathbf{k},\downarrow}$ terms (see Sec. V), we obtain as different components of the BE-AHC:

$$J_1^x = e\int d\mathbf{k}\, v^x_{sj,k,\uparrow}\left(\rho^{(E^1,B)}_{\mathbf{k},\uparrow} + \rho^{(E^1,B)}_{-\mathbf{k},\downarrow}\right) = \frac{e^2}{\hbar}\frac{\omega_B E}{(s^2 + \omega_B^2)}\int d\mathbf{k}\, v^x_{sj,k,\uparrow}\left\{\partial_y + \frac{\omega_B}{s}\partial_z\right\}\left(\rho^{(0)}_{\mathbf{k},\uparrow} - \rho^{(0)}_{-\mathbf{k},\downarrow}\right),$$

$$J_2^x = -\frac{e^2}{\hbar}\int d\mathbf{k}\left(\Omega_{k,\uparrow}^{y} v_{0,k,\uparrow}^{y} + \Omega_{k,\uparrow}^{z} v_{0,k,\uparrow}^{z}\right) B\left(\rho_{\mathbf{k},\uparrow}^{(E^1,B)} + \rho_{\mathbf{k},\uparrow}^{(E^1)} + \rho_{-\mathbf{k},\downarrow}^{(E^1,B)} + \rho_{-\mathbf{k},\downarrow}^{(E^1)}\right) =$$
$$-\frac{e^3}{\hbar^2}\frac{EB}{(s^2+\omega_B^2)}\int d\mathbf{k}\left(\Omega_{k,\uparrow}^{y} v_{0,k,\uparrow}^{y} + \Omega_{k,\uparrow}^{z} v_{0,k,\uparrow}^{z}\right)\left(\omega_B \partial_y - s\partial_z\right)\left(\rho_{\mathbf{k},\uparrow}^{(0)} - \rho_{-\mathbf{k},\downarrow}^{(0)}\right),$$

$$J_3^x = e\int d\mathbf{k}\, v_{0,k,\uparrow}^{x}\left(\rho_{\mathbf{k},\uparrow}^{(E^1,B)} - \rho_{-\mathbf{k},\downarrow}^{(E^1,B)}\right) = \frac{e^2}{\hbar}E\int d\mathbf{k}\, v_{0,k,\uparrow}^{x}\left\{\frac{e}{\hbar}\frac{\omega_B \Omega_{k,\uparrow}^{x} B}{(s^2+\omega_B^2)^2}\left(2s^2\partial_y +\right.\right.$$
$$\left.\frac{\omega_B(3s^2+\omega_B^2)}{s}\partial_z\right) + \frac{\omega_B v_{sj,k,\uparrow}^{z}}{(s^2+\omega_B^2)}\left(\left(\frac{\partial_z}{v_{0,k,\uparrow}^{y}} - \frac{\partial_y}{v_{0,k,\uparrow}^{z}}\right) - \frac{\omega_B}{s}\left(\frac{\partial_y}{v_{0,k,\uparrow}^{y}} + \frac{\partial_z}{v_{0,k,\uparrow}^{z}}\right)\right) - \left.\frac{k^z}{k}\frac{w_a B}{s^2}\partial_x\right\}\left(\rho_{\mathbf{k},\uparrow}^{(0)} - \rho_{-\mathbf{k},\downarrow}^{(0)}\right),$$

$$J_4^x = \frac{e^2}{\hbar}E\int d\mathbf{k}\Omega_{k,\uparrow}^{y}\left(\rho_{\mathbf{k},\uparrow}^{(B)} - \rho_{-\mathbf{k},\downarrow}^{(B)}\right) = \frac{e^3}{\hbar^2}\frac{EB}{(s^2+\omega_B^2)}\int d\mathbf{k}\Omega_{k,\uparrow}^{y}\left\{s\left(v_{0,k,\uparrow}^{y}\partial_z - v_{0,k,\uparrow}^{z}\partial_y\right) -\right.$$
$$\left.\omega_B\left(v_{0,k,\uparrow}^{z}\partial_z + v_{0,k,\uparrow}^{y}\partial_y\right)\right\}\left(\rho_{\mathbf{k},\uparrow}^{(0)} - \rho_{-\mathbf{k},\downarrow}^{(0)}\right).$$

For a simple parabolic and spin-degenerate bandstructure, many terms in the above equations do not survive the k-space integration and the equations get simplified to:

$$J_1^x = \frac{e^3}{\hbar^2}\frac{\gamma EB}{(s^2+\omega_B^2)}\int d\mathbf{k}\left\{\Omega_{k,\uparrow}^{z} v_{0,k,\uparrow}^{y}\partial_y - \Omega_{k,\uparrow}^{y} v_{0,k,\uparrow}^{z}\frac{\omega_B}{s}\partial_z\right\}\left(\rho_{\mathbf{k},\uparrow}^{(0)} - \rho_{-\mathbf{k},\downarrow}^{(0)}\right),$$

$$J_2^x = -\frac{e^3}{\hbar^2}\frac{EB}{(s^2+\omega_B^2)}\int d\mathbf{k}\left(\omega_B\Omega_{k,\uparrow}^{y} v_{0,k,\uparrow}^{y}\partial_y - s\Omega_{k,\uparrow}^{z} v_{0,k,\uparrow}^{z}\partial_z\right)\left(\rho_{\mathbf{k},\uparrow}^{(0)} - \rho_{-\mathbf{k},\downarrow}^{(0)}\right),$$

$$J_3^x = -\frac{e^2}{\hbar}\frac{\omega_B^2 E}{s(s^2+\omega_B^2)}\int d\mathbf{k}\, v_{0,k,\uparrow}^{x} v_{sj,k,\uparrow}^{z}\left(\frac{\partial_y}{v_{0,k,\uparrow}^{y}} + \frac{\partial_z}{v_{0,k,\uparrow}^{z}}\right)\left(\rho_{\mathbf{k},\uparrow}^{(0)} - \rho_{-\mathbf{k},\downarrow}^{(0)}\right),$$

$$J_4^x = -\frac{e^3}{\hbar^2}\frac{EB\omega_B}{(s^2+\omega_B^2)}\int d\mathbf{k}\Omega_{k,\uparrow}^{y}\left(v_{0,k,\uparrow}^{z}\partial_z + v_{0,k,\uparrow}^{y}\partial_y\right)\left(\rho_{\mathbf{k},\uparrow}^{(0)} - \rho_{-\mathbf{k},\downarrow}^{(0)}\right).$$

For these simplifications we have considered the following relations: $\bar{g}_{c,\mathbf{k},\uparrow} = \bar{g}_{c,-\mathbf{k},\uparrow}$, $\boldsymbol{\Omega}_{k,\uparrow} = \boldsymbol{\Omega}_{-k,\uparrow}$, $\boldsymbol{v}_{0,k,\uparrow} = -\boldsymbol{v}_{0,-k,\uparrow}$, and $\boldsymbol{v}_{\boldsymbol{sj},k,\uparrow} = \sum_{k'} w_{k'k}\boldsymbol{\Omega}\times(\mathbf{k}' - \mathbf{k}) = -\boldsymbol{\Omega}\times\sum_{k'} w_{k'k}(k - k'\cos(\theta))\widehat{\boldsymbol{k}} = \gamma\mathbf{k}\times\boldsymbol{\Omega} = \frac{m}{\hbar}\gamma\boldsymbol{v}_{0,k,\uparrow}\times\boldsymbol{\Omega}$. In the simplified expression for $\boldsymbol{v}_{\boldsymbol{sj},k,\uparrow}$, we replaced $\mathbf{k}$ with $\frac{m}{\hbar}\boldsymbol{v}_{0,k,\uparrow}$, which is plausible for parabolic bands with no spin-splitting. This replacement also makes it easy to visualize the symmetry relation $\boldsymbol{v}_{\boldsymbol{sj},k,\uparrow} = \boldsymbol{v}_{\boldsymbol{sj},-k,\downarrow}$. The skew scattering term vanishes if we ignore any k-dependence of $w_a$, therefore we ignore it by considering it very weak compared to the side-jump and Berry curvature contributions [10]. We also note that $\boldsymbol{\Omega}_{k,\uparrow}$ is a non-abelian Berry curvature, i.e., a Berry curvature in the presence of spin-degeneracy, and does not behave like a magnetic monopole in k-space. Such Berry curvatures are even-functions of k for the simple case of zero spin splitting. In contrast, the abelian Berry curvatures are odd functions of k, reflect spinless bandstructures, and act as magnetic monopoles in k-space.

In the above equations for $J^x$ we still have terms which do not depend on the sign of the external magnetic field. We drop these terms, since in our measurements we specifically extracted the current components which switch their directions upon inverting the direction of the magnetic field. We now rewrite the BE-AHC components according to the originating mechanisms:

$$J_{sj}^x = J_1^x = \frac{e^3}{\hbar}\frac{2\gamma BEI(\omega)}{(s^2+\omega_B^2)(i\omega+w^{\mathrm{sf}})}\int d\mathbf{k}\,\Omega_{k,\uparrow}^{z}\left(v_{0,k,\uparrow}^{y}\right)^2\frac{d\bar{g}_{c,\mathbf{k},\uparrow}}{d\epsilon_{\mathbf{k}}}, \tag{25a}$$

$$J^x_{BC} = J^x_2 = \frac{e^3}{\hbar}\frac{2sBEI(\omega)}{(s^2+\omega_B^2)(i\omega+w^{\mathrm{sf}})}\int d\mathbf{k}\,\Omega^z_{k,\uparrow}\left(v^z_{0,k,\uparrow}\right)^2\frac{d\bar{g}_{c,\mathbf{k},\uparrow}}{d\epsilon_{\mathbf{k}}}\,, \tag{25b}$$

with $J^x_{sj}$ and $J^x_{BC}$ denoting the side-jump and the Berry-curvature contributions to the BE-AHC, respectively. As can be seen from equations (25), the time and frequency dependences of the two contributions differ from each other. We note that the above calculations are done under the approximation of small scattering angles. While the calculations yield the time responses of the side-jump and Berry-curvature contributions to the BE-AHC, they cannot be used to compare the absolute strength of the contributions with each other.

## VIII. Expressions for HCs

In this section we will obtain the final expression for the HC. Inserting equation (24) into equation (17b) and, consecutively, equation (17b) into equation (16b) with omission of the $\rho^{(0)}_{\mathbf{k},\uparrow} - \rho^{(0)}_{-\mathbf{k},\downarrow}$ terms (see Sec. V), and additionally omitting the term which does not switch it's sign upon inversion of the magnetic field direction, we obtain:

$$J^y_{HC} = e\int d\boldsymbol{k}\,v^y_{0,k,\uparrow}\left(\rho^{(E^1,B)}_{\mathbf{k},\uparrow} - \rho^{(E^1,B)}_{-\mathbf{k},\downarrow}\right) = \frac{e^2}{\hbar}\frac{\omega_B E}{(s^2+\omega_B^2)}\int d\boldsymbol{k}\,v^y_{0,k,\uparrow}\partial_y\left(\rho^{(0)}_{\mathbf{k},\uparrow} + \rho^{(0)}_{-\mathbf{k},\downarrow}\right)$$
$$= \frac{2e^2\omega_B EI(\omega)}{(s^2+\omega_B^2)(i\omega+w^{\mathrm{E}})}\int d\boldsymbol{k}\left(v^y_{0,k,\uparrow}\right)^2\frac{dg_{l,\mathbf{k},\uparrow}}{d\epsilon_{\mathbf{k}}}\,, \tag{26}$$

Here we ignore the $\bar{g}_{l,\mathbf{k},\uparrow}$ term considering it very weak compared to the term involving $g_{l,\mathbf{k},\uparrow}$. Comparing equation (26) to equations (25) we see that the time and frequency responses of the HC and BE-AHCs differ from each other. This will be further analyzed in the next section.

## IX. THz responses from valance and conduction bands

Examining equation (26) reveals that the amplitude of $J^y_{HC}$ is proportional to $\gamma^{-2}$. Since in the VB $\gamma$ is much larger as compared to the CB, the CB contribution to the HC dominates and the VB contribution to the HC can be ignored. This allows us to write the THz spectral response of the total HC as:

$$S^y_{HC}(\omega) \propto \frac{i\omega BI(\omega)}{(s^2+\omega_B^2)(i\omega+w^{\mathrm{E}})}G(\omega)\,, \tag{27}$$

with $\gamma$ and $\omega_B$ being the CB momentum scattering rate and cyclotron frequency respectively. $G(\omega)$ is a filter function accounting for THz propagation and detection effects. The $i\omega$ in the numerator of equation (27) results from the differentiation of the current required to obtain the electric THz field.

In contrast to the HCs, the contribution from the VB cannot be ignored for the BE-AHCs for two reasons: (i) the BE-AHC is only proportional to $\gamma^{-1}$ and (ii) the anomalous coefficients in the VB are much stronger than their CB counterparts [11]. Moreover, we only consider the side-jump contribution to the BE-AHC, as in GaAs it is supposed to be much stronger than the Berry curvature contribution. Therefore, for the BE-AHC the THz spectral response is written as:

$$S_{AHC}^{x}(\omega) \propto i\omega B I(\omega)\left\{\frac{\gamma_{cb}}{(s_{cb}^{2}+\omega_{B}^{2})(i\omega+w_{cb}^{\mathrm{sf}})} + C\,\frac{1}{\gamma_{vb} w_{vb}^{\mathrm{sf}}}\right\} G(\omega)\ , \qquad (28)$$

where the subscripts cb and vb represent CB and VB parameters, respectively. $C$ is a constant accounting for various microscopic parameters, whose calculation is out of scope of this supplementary material. We ignore the cyclotron frequencies from the valance band, since they are too small to be observed within our THz spectral windows and, more importantly, the cyclotron effects in the VB are heavily suppressed due to the very large $\gamma_{vb}$.

Analyzing the spectral responses expressed through equations (27) and (28), we can state the following. For HCs, the THz spectra have peaks corresponding to the CB cyclotron frequency. In contrast, the CB cyclotron motion leads to a dip in the BE-AHC THz spectra. This spectral dip cannot be explained without a significant contribution from the VB.

Another feature appeared in the experimental BE-AHC studies presented in the main article. The BE-AHC appears to be inverted for certain excitation conditions (low magnetic fields and low photon excitation energies). This leads to the possibility that $J_{sj}^{x}$ and $J_{BC}^{x}$ are inverted to each other and each of the two contributions dominates for different excitation conditions. Although a difference between the current responses of $J_{sj}^{x}$ and $J_{BC}^{x}$ can be explained with the relation $J_{BC}^{x} \propto \left(J_{sj}^{x} + \frac{i\omega}{\gamma} J_{sj}^{x}\right)$, a complete current inversion is not apparent from our simple calculations of $J_{sj}^{x}$ and $J_{BC}^{x}$, see equations (25). Most likely, this current inversion has its origin in bandmixing of continuum states. Such current inversion resulting from bandmixing has already been observed in earlier studies of injection and shift currents [12,13].